\documentclass[12pt,preprint]{aastex}

\slugcomment{\today}

\begin{document}

\title{Low-Mass Tertiary Companions to Spectroscopic Binaries I: Common Proper Motion Survey for Wide Companions using 2MASS}

\author{Peter R.\ Allen\footnote{Visiting Astronomer, Kitt Peak National Observatory and Cerro Tololo InterAmerican Observatory, National Optical Astronomy Observatory, which is operated by the Association of Universities for Research in Astronomy (AURA) under cooperative agreement with the National Science Foundation.}}
\affil{Department of Physics and Astronomy, PO Box 3003, Franklin and Marshall College, Lancaster, PA 17604; peter.allen@fandm.edu}

\author{Adam J.\ Burgasser\footnote{Visiting Astronomer at the Infrared Telescope Facility, which is operated by the University of Hawaii under Cooperative Agreement no. NCC 5-538 with the National Aeronautics and Space Administration, Science Mission Directorate, Planetary Astronomy Program.}}
\affil{Department of Physics, University of California, San Diego, CA 92093}

\author{Jacqueline K. Faherty${^2}$}
\affil{Department of Astrophysics, American Museum of Natural History, Central Park West at 79th Street, New York, NY 10034}

\author{J.\ Davy Kirkpatrick${^2}$}
\affil{IPAC, California Institute of Technology, Pasadena, CA 91125}

\begin{abstract}

We report the first results of a multi-epoch search for wide (separations greater than a few tens of AU), low-mass tertiary companions of a volume-limited sample of 118 known spectroscopic binaries within 30~pc of the Sun, using the 2MASS Point Source Catalog and follow-up observations with the KPNO and CTIO 4m telescopes.  Note that this sample is not volume-complete but volume-limited, and, thus, there is incompleteness in our reported companion rates.  We are sensitive to common proper motion companions with separations from roughly 200~AU to 10,000~AU ($\sim10\arcsec \rightarrow~\sim10\arcmin$).  From 77 sources followed-up to date, we recover 11 previously known tertiaries, three previously known candidate tertiaries, of which two are spectroscopically confirmed and one rejected, and three new candidates, of which two are confirmed and one rejected.  This yields an estimated wide tertiary fraction of $19.5^{+5.2}_{-3.7}\%$.  This observed fraction is consistent with predictions set out in star formation simulations where the fraction of wide, low-mass companions to spectroscopic binaries is $>$10\%, and is roughly twice the wide companion rate of single stars.

\end{abstract}

\section{Introduction}

Formation simulations have had a difficult time modeling the very close separations of many spectroscopic binaries.  A mechanism is needed to draw angular momentum away from an already close pair of objects \citep{kis98}.  Recent star formation simulations \citep{sd03,dd04,umb05} show that one potential mechanism for the transfer of angular momentum is through three-body interactions.  The third bodies used, in these cases, are cool dwarfs.  Cool dwarfs are stellar and sub-stellar objects with spectral types $\gtrsim$ M and masses less than a few tenths of a solar mass.  Interactions between loosely bound or totally unbound low-mass objects can dramatically tighten already-close orbits.  These results predict that spectroscopic binary systems should have a larger fraction of wide cool companions than so-called `single' stars.  

The dynamic interaction simulations of \citet{sd03} and \citet{dd04} produce some testable predictions.  The simulations of \citet{dd04} predict that, if a cool dwarf is found to be in a stable, $>$10~AU orbit, its primary is frequently ($\sim75\%$) a tight spectroscopic binary.  A similar qualitative result is found in the dynamic simulations of \citet{sd03}.  Recent work by \citet{law10}, which studied a sample of known, very wide M dwarf binary systems, found that $45^{+18}_{-16}\%$ were higher order multiples.  This supports the simulation predictions.  Note that `wide' in this work means separations $\gtrsim 10$'s of AU.

The Delgado-Donate et~al.\ simulations found that $\sim$10\% of their tight multiple systems survive with wide, cool dwarf companions by the end of their simulations, 10.5~Myr.  There are many empirical measurements of the cool dwarf companion frequency to stars.  \citet{gizis01} estimated a wide cool dwarf companion frequency of $18\%{\pm}14\%$, based on only three L and T dwarf companions.  \citet{jc09} examined 21 FGK stars within 20~pc, and found no companions in a range of 20~AU - 250~AU down to masses of 50~$M_J$.  \citet{pl05} used the Hubble Space Telescope to examine 45 young stars ($\sim$0.15~Gyr) at separations from 15~AU to 200~AU and masses down to typically 30~$M_J$ to find or confirm 8 cool companions.  \citet{mz04} examined a sample of $\sim$300 single, G, K, and M stars at separations from 75 AU to 1200 AU and masses as low as $\sim$5~$M_J$ and found a cool dwarf binary frequency of 1-2\%.  \citet{mh04,mh06,mh09} examined $\sim$250 `Solar Analogs' (F5-K5 stars).  They found two new brown dwarf and 24 new stellar companions with separations from 28~AU to 1590~AU and probed masses down to ${\sim}10~M_J$.  They calculate an ultracool companion frequency of $\sim3\%$, which is marginally higher than that of \citet{mz04}.  

All of the above works are consistent with a low, wide, cool dwarf companion fraction of only a few percent.  However, the primaries are all apparently single stars, which does not test the simulation results on tertiary companions.  \citet{tok06} conducted such a study of 165 spectroscopic binaries, in which they searched for wide companions using the 2MASS database.  A subset of those objects (62) were observed at high spatial resolution with NACO on the VLT.  They found a very high tertiary rate, adjusted for incompleteness, of 63\%.  They also found that the fraction of spectroscopic binaries with tertiary components is a strong function of spectroscopic binary period.  Those with very short periods (less than 12 days) almost all have wide companions (96\%).

It should be noted that other groups have conducted common proper motion (CPM) comparisons between various wide-field surveys, though none has specifically targeted wide companions to spectroscopic binaries.  The Brown Dwarf Kinematics Project \citep{jf09,jf10} has studied the kinematics of ultra cool dwarfs and found or confirmed several wide cool companions to stellar primaries, including one spectroscopic binary.  \citet{lb07} performed a re-analysis of the Digitized Sky Survey using the custom built software package SUPERBLINK.  They found that ${\sim}9.5\%$ of Hipparcos stars have companions at separations wider than 1000 AU and proper motions great than $0{\farcs}15/yr$.  There have been several papers cross-correlating 2MASS and SDSS which have reported individual discoveries, such as \citet{met08} and \citet{kg11}.  There have also been large systematic cross-correlations, such as \citet{jdk10} and \citet{sc09}, 2MASS to SDSS; or \citet{nd09}, 2MASS to UKIDSS.  However, these works focused on finding individual field objects with large proper motions, not finding multiple systems.  Finally there is the Slowpokes survey \citep{slowpokes} which cross-correlates USNO-B to SDSS and was designed to look for CPM companions.  They focused on the general field population and only searched separations $\le 180"$.  This is narrower than our search radius and is limited to the optical and will not have the same sensitivity to extremely cool objects as our near-infrared survey.

Here we report the first results from a study to measure the wide tertiary fraction around spectroscopic binaries via CPM, using 2MASS as a first epoch and our own deep near-infrared imaging as the second epoch, which \citet{tok06} did not carry out.  Section 2 describes the experimental setup and sample selection, while Section 3 outlines the wide-field NIR imaging campaign and the data reduction procedures.  Sections 4 and 5 detail our CPM analysis techniques and discusses the results, respectively.  Section 6 summarizes the current work and our results.

\section{Experimental Design and Sample Selection}

We required our sample to differentiate statistically between the $1-2\%$ observed wide companion rate of `single' stars \citep{mz04} and the $>10\%$ results predicted by \citet{sd03}, \citet{dd04}, and \citet{umb05}.  Assuming Poisson statistics, a $1\%-2\%$ frequency can be distinguished from a 10\% frequency at the 3$\sigma$ level when $\sim$100 objects are observed.  Our spectroscopic binary sample of $\sim$100 was selected to detect bright to very faint wide companions using a CPM method.

We used the 2MASS Point Source Catalog, which has astrometric uncertainties of ${\sim}0{\farcs}2$ and which was taken between 1997 and 2001, to provide the first epoch.   Second epoch near-infrared imaging was conducted with the KPNO 4m Flamingos \citep{flam} and CTIO 4m ISPI \citep{ispi1,ispi2} instruments.  These cameras have plate scales of $0{\farcs}3156$/pix and $0{\farcs}3$/pix respectively.  Given our previous experience at measuring astrometry, we expected to be able to measure the position of objects in our fields to within ${\sim}10\%$ of the typical seeing measurements.  So, with the average seeing at KPNO and CTIO of $\sim1\farcs0$, we expected to obtain astrometry good to $\sim0\farcs1$.  For a robust CPM measurement, we chose an approximately $5\sigma$ shift in position to reject chance alignments with background objects whose motion closely matches that of the primary's.  This is particularly relevant for those spectroscopic binaries whose motion is relatively small because our measurement uncertainties and systematics can make another, slower-moving background object appear to be a genuine CPM companion.  A $\sim5\sigma$ detection of CPM therefore requires a minimum motion of $0\farcs1$/yr.

We also wanted our data to be sensitive to objects fainter than 2MASS by two to three magnitudes (similar sensitivities as UKIDSS \citet{ukidss}).  In this way we would be able to use the data reported here as a first epoch of deep observations sensitive to brown dwarfs considerably cooler than known T dwarfs and, potentially, the long sought after, and recently discovered, new spectral type range of Y dwarfs \citep{davy08,cush11}.  As stated on the Flamingos Website, the ISPI exposure time calculator may be used for both instruments.  The calculator indicates that a 15 minute exposure time yields $5\sigma$ detection limits of 20.1, 19.5, and 18.8 at J, H, and K$_s$ respectively.  These detection limits also represent significantly lower mass limits, for typical Galactic disk ages of a few Gyr, from our observations (${\sim}10-20$ M$_J$) than is possible with 2MASS alone (${\sim}50-60$ M$_J$).  Also note that \citet{jf09} has found that the typical proper motion of $\ge$M7 dwarfs with distances around 15 pc $-$ 30 pc is ${\ge}0{\farcs}2/yr$.  This motion is well above the limit of our survey and the distance range corresponds to objects we could easily detect in our images.  Thus, we will also be able to select for fast-moving objects in our fields to search for nearby very cool objects.

The sample was selected from `The Ninth Catalogue of Spectroscopic Binaries' \citep{sb9} (SB9).  This catalog is a compilation of known binaries from the literature and comes with a few significant biases.  When the SB9 catalog was first compiled (2004), there were about 1200 potential new systems in the literature that had not yet been evaluated for inclusion. Additionally, the surveys for spectroscopic binaries that form the core of the SB9 catalog are biased against fainter, lower-mass binary systems. \citet{sb9} indicates that this bias begins around K spectral types.  Thus, while we treat this sample as a whole, it does not necessarily reflect the statistics in a volume-limited sample which would be dominated by K and M dwarfs \citep{jb2011}.

We ran the full catalogue of 2386 binaries through the Vizier interface of the Hipparcos catalog to determine the distance and proper motion of each object.  Given the criteria derived at the beginning of this section, a volume-limit of 30~pc yielded an optimal sample size of 118 targets with appropriate motions (see Figure \ref{fig:pmvd}, left-hand panel), which are listed in Table \ref{tab:targs}.  By choosing nearby targets, we are sensitive to some of the brightest, lowest-mass brown dwarfs in the sky.  Our data are able to find companions with separations greater than $\sim$10\arcsec.  This corresponds to projected separations greater than $\sim$200~AU and periods greater than 1000 years for the typical primary distances and masses within our distance limit of 30~pc.

The resultant sample is both volume and proper-motion limited to those sources in the Hipparcos database.  This introduces some biases, including kinematics: only objects moving faster than $\sim14$~km/s at 30~pc were selected.  This kinematic bias favors the selection of older, faster moving systems.  Thus, we are biased against detecting the youngest systems with the lowest-mass companions, as they will have cooled to temperatures below our detection limit.  However, without knowing {\it a priori} what the ratio of young systems to old systems is, this bias is very difficult to estimate.  There is also a spectral type bias against M dwarf primaries, due to the V magnitude limit of Hipparcos and the relative lack of data on M dwarf spectroscopic binaries.  Also, note that this sample is {\it not} volume-complete.  As can be seen in the right hand panel of Figure \ref{fig:pmvd}, the number of objects continues to increase to distances of roughly 20~pc and then levels off.  This is likely a result of the magnitude limit of the Hipparcos catalog.  We are biased again the fainter, lower mass, more distant primaries.  These biases will be quantitatively accounted for via a Bayesian/Monte Carlo simulation in a future paper.

Figure \ref{fig:sptdist} displays two histograms of the spectral type distribution of our primary sample.  The left-hand panel of Figure \ref{fig:sptdist} shows the overall spectral type distribution, and we can see that we have mostly F, G, and K stars.  Thus, our sample will most accurately examine the tertiary fraction of spectroscopic binaries composed of F, G, and K stars.  The right-hand panel of Figure \ref{fig:sptdist} shows how the spectral type distribution changes as a function of distance, which demonstrates that our sample is magnitude limited.  This makes sense, given the selection biases of the SB9 catalog towards brighter, more distant, earlier spectral types.  Thus, within our sample, we are likely missing G and K-type spectroscopic binaries in our larger distance bins.  Given that we have a comparable overall number of F, G, and K stars in the sample, we are probing the statistical companion rate as a function of spectral type equally across those types.

Another factor to be taken into consideration is that of orbital motion.  We expect to locate companions as close as $10\arcsec - 20\arcsec$ from the central binaries.  Given our smallest target distance of around 5 pc, the minimum angular separation of $10\arcsec$, and the typical time baselines of 10 years, the orbital motion is maximal at ${\sim}120$ mas/yr.  This is on the same order as our typical 2$\sigma$ CPM search criterion limit we derive in Section \ref{sec:cpmtech} of $\sim100$ mas/yr.  However, we only have a handful of systems at distances less than $\sim$10 pc that would have such large maximal orbital motions.  This will be discussed in more detail in Section \ref{sec:cpmtech}, but we conclude that the orbital motion effects will be minimal.

\section{Observations and Data Reduction}

\subsection{Observations}

The observations described here were carried out during four successful observing runs: May 3-6 2007 (KPNO); June 3-6 2007 (CTIO); January 20-23 2008 (CTIO); January 28-30 2008 (KPNO).  Table \ref{tab:targs} lists when and where each target was observed.

We obtained standard calibration data in the afternoon before each night's observing, including sets of dome flats in all three bands ($JHK_s$), bias frames, and dark frames.  Our science data consisted of dithered $JHK_s$ imaging of each target field.  Dithering allowed the construction of accurate maps of the IR background in each exposure which were then subtracted from each individual frame.  We set individual dither position exposure times to keep the background well below the beginning of the non-linear regime ($\sim$10000-13000 ADU for both instruments).  We could not avoid saturating on the very bright central spectroscopic binaries.  The saturated areas typically effect the inner part of the image to radii of 10 - 15 arcseconds but could reach to much larger radii (see section \ref{sec:sens}).

Using IPSI, we modified a 15-point random dither script for each band pass and adjusted the exposure time and number of coadds according to the background at that band pass each night, such that the total exposure time per pointing was 60 seconds.  For Flamingos the procedure was slightly different, due to the lack of a coadd feature.  We instead repeated our dither patterns, 5X5 for the May 2007 run and 4X4 for the January 2008 run.  We selected appropriate exposure times to keep the background in the linear regime.  This required us to repeat the dither pattern two to four times.

\subsection{Data Reduction}

\subsubsection{Basic Reduction}

Data reduction procedures for both instruments were performed identically, using IRAF tools developed for ISPI \footnote{http://www.ctio.noao.edu/instruments/ir\_instruments/ispi/New/UsersGuide/datared.html} and the publically available software packages WCSTOOLS \citep{mink} and SWarp.  For each night of data, we first created a bad pixel mask, usually from a dome flat image, using the IRAF task {\it ccdmask}.  This mask was applied to all data for that night, including calibration data, using the IRAF task {\it fixpix}.  This task linearly interpolates over the marked bad pixels from {\it ccdmask} using nearby `good' pixels.  We reduced the $JHK_s$ data for each science target separately in the following manner.  First, we dark-subtracted the raw frames and corrected for non-linearity.  Next, using the {\it xdimsum} IRAF package, we estimated the sky background, subtracted it, and masked the holes (the result of subtracting dithered images with stars in them from one another).  In the final step, we flat-fielded the sky and dark-subtracted images.

\subsubsection{Mosaiking and Coordinate Solutions}
\label{sec:coord}

Since the main goal of this project was to obtain accurate proper motions for the objects detected in our fields, we needed to combine the individual frames into a final image and refine the resultant coordinate solution carefully.  We first oriented each final reduced image, such that North is up and East is left.  Sources were then selected to create a rough coordinate solution.  This was done using the {\it starfind} IRAF task.  We selected these initial lists to be unsaturated sources in each field of medium brightness, typically from $10 \rightarrow 14$ mag at $J$-band, to match against the 2MASS PSC from the full release \citep{2mass}.  We then fed the output from {\it starfind} into the WCSTOOLS task {\it imwcs}, which provided a preliminary coordinate solution.  

Next, we refined the rough coordinate solution provided by {\it imwcs}, and corrected distortion across the field of view using the IRAF task {\it ccmap}, which we ran interactively for all images via the {\it cirred} package task {\it do\_ccmap}.  This was done to determine the average distortion solutions across all of the images.  In {\it ccmap}, distortion was quantified by polynomial fit in x and y pixel positions, with orders set to either four, for relatively sparse fields (those with fewer than $\sim$40 sources, which corresponds to roughly a third of our observed sample), and six, for more populated fields.  When {\it ccmap} was run on each image, we interactively removed sources that have large differences from their 2MASS coordinates until the residuals of the coordinate solutions were on the order of $0{\farcs}1$.  The number of sources removed varied from image to image and could be as few as 4 or 5 or as many as 30 or 40 for very crowded fields.  This was never more that ${\sim}10\%$ of the sources in any given field.  The removed sources were typically mismatched artifacts due to the regions around the bright primary.  Note that the residuals we obtained were similar to those of the 2MASS PSC and matched our expected astrometric precision.  Also, similar coordinate solutions and final coordinates were found in all three bands, $J$, $H$, and $K_s$.  Finally, we used SWarp, including the average distortion correction created via {\it do\_ccmap}, to perform the final mosaiking.

\section{Identification of CPM Companions}

\subsection{Technique}
\label{sec:cpmtech}

We measured our second epoch astrometry from the $J$-band mosaicked images for the majority of our sources.  Again, we selected sources using the IRAF task {\it starfind}, extracted photometry using {\it phot}, and measured RA and DEC coordinates from the image WCS in the fits header, using {\it wcsctran}.  We manually cleaned this catalog of sources of all spurious objects by visual inspection, using ds9.  The majority of these spurious sources were associated with the bright halo of the saturated central spectroscopic binary and typically numbered between a few dozen and several hundred per field.  We then compared the cleaned list of sources to objects found in the field selected from the 2MASS PSC \citep{2mass} using custom-built Fortran scripts.  

Comparison of our astrometry to that from 2MASS was a two-fold process.  The first run looked for objects that did not move (defined as objects in our data that are within $0\farcs5$ of their 2MASS positions) and marked them as matched.  This helped to prevent mismatches with nearby faint sources when looking for moving objects.  The second run ignored these matched, non-moving objects and looked for objects with motions of a similar magnitude to that of the target spectroscopic binary in the field.  Our program then filtered the list of `moving' objects further by determining the uncertainties on all the proper motion measurements for that field.  The majority of objects in a given field display little to no motion, and those measurements follow a normal distribution.  Thus, we can use the standard deviation of that normal distribution as an estimate of the uncertainty of our proper motion measurement.  We used the $2\sigma$ limit as the threshold to determine the proper motion uncertainty in both ${\mu}_{\alpha}$ and ${\mu}_{\delta}$ in mas/yr.  The limits ran from 62 mas/yr up to 160 mas/yr with an average value of ${\sim}$100 mas/yr.  These limits naturally include the $\sim0\farcs1$ residuals from the coordinate solution.  We considered any object whose motion falls in that 2$\sigma$ to be a CPM companion candidate.  Example plots of the CPM diagrams for fields both with and without a companion can be seen in Figure \ref{fig:cpm}.

As mentioned in Section 2, the orbital motion of candidate companions must be examined.  We determined the maximum orbital motion of a companion in all of our fields by taking the typical inner working angle of $\sim$10$\arcsec$, the distance to each primary, and typical primary masses of $\sim$2~M$_{\odot}$ and applying Kepler's 3rd Law.   These maximal values were then compared to the 2$\sigma$ proper motion uncertainty values.  In all cases, save one, the maximum orbital motion was less than the uncertainty used to select candidates.  Only in the field of HIP 88601, our closest observed target, was the maximal motion of 123 mas/yr greater the proper motion uncertainty of 120 mas/yr.  Additionally, several of our targets that fall within 10 pc also have maximal orbital motions that are within a factor of 2 of the proper motion uncertainties.  However, at radii of 15$\arcsec$ to 20$\arcsec$ the orbital motion falls well below the uncertainties.  These fields represent about 10\% of our observed sample, but it is only a relatively small fraction of each field in which we could have thrown out a potential candidate.  Thus, we most likely did not remove any candidates because of their orbital motion.

Despite the above selection thresholds, mismatches can still occur, which are typified in the left-hand panel of Figure \ref{fig:cpm}.  To remove this last set of spurious sources, we employ one further candidate selection criterion, visual comparison of the Flamingos/ISPI images with the Digitized POSS plates.  Given the very red optical-NIR colors of brown dwarfs, we do not expect them to have counterparts in the POSS R-band.  Nearly all of the candidates that passed the earlier criteria we found to be spurious.  They typically occurred in fields around primaries with higher proper motions ($>{\sim}0{\farcs}4/yr$), as a result of mismatches between brighter 2MASS sources with small motions (hence not thrown out in the first sweep) and faint sources from our Flamingos/ISPI images.  These non-moving, faint sources in our data were matched to brighter 2MASS sources because they are too faint to appear in the 2MASS PSC.  This leads the matching software to find the closest available unmatched 2MASS source, which is always a nearby bright source.  These sources are also typically matched against a faint source in the Digitized POSS, and the bright 2MASS source has a corresponding bright source in our Flamingos/ISPI data.  So, we did not discard any potentially very red companions.

Through the visual comparison of each candidate to the DPOSS R-band plates we reduced the list of candidates from dozens to 13, which are listed in Table \ref{tab:cand} along with five other known bright companions that are saturated in our data and thus not detected in this analysis.  We uncovered the five saturated companions via a literature search using the SIMBAD database for each of the target spectroscopic binaries we observed.  These companions are all very bright, with apparent J magnitudes less than 8.  For those known objects whose motion we measured, we successfully recovered their known proper motions within our uncertainties.  This was a final verification step for the coordinate solution process discussed in Section \ref{sec:coord}.

\subsection{Candidate Follow-up}

Among our CPM companions, we identified 14 that had been previously noted as verified or candidate companions in the literature.  Of these, 11 are confirmed companions, discovery references are given in Table \ref{tab:cand}, based on their motions and spectral types, the latter providing a spectrophotometric distance that is consistent with the Hipparcos distance of the spectroscopic binary.  Three additional candidate companions were also recovered (HIP 41211C \citep{tok06}; HIP 83608C \citep{ccdm2}; HIP 101769C \citep{ccdm2}), for which CPM or spectroscopic confirmation had not yet been obtained.  Finally, we identified three new candidate companions.  

Candidates were followed up with low-resolution $1-2.5~\micron$ spectroscopy using the prism setting of Spex \citep{spex} on NASA's InfraRed Telescope Facility (IRTF).  In all of the observations the spectrograph slit was aligned with the parallactic angle to minimize the effects of differential refraction.  The data were taken in the ABBA dither pattern format to allow for easy subtraction of the infrared background.  We reduced these data using Spextool \citep{spextool}, an IDL based software package designed specifically for SpeX data.  This reduction package removes the background, traces and extracts the spectral data, and wavelength and flux-calibrates the data with an A0V standard star.  The final reduced and calibrated spectra are then compared to spectral standards from the IRTF Spectral Library \citep{cush05,rayn08}, and are displayed in Figures \ref{fig:hip41211} - \ref{fig:hip97944}.

\subsubsection{Previously Identified Candidate Companions}

{\noindent {\it HIP 41211C:} The closer companion to HIP 41211, in Table \ref{tab:cand}, was noted as a visual companion in \citet{tok06}.  Our data demonstrated CPM, and our SpeX prism spectrum yielded a spectral type of M4.5 (Figure \ref{fig:hip41211}), which is consistent with being a physical companion.  This confirmation, in addition to the very wide M5.5 companion to HIP 41211 found by \citet{inr03}, makes this system the only confirmed quadruple currently known in this sample.  The projected separation for the newly confirmed companion is 980~AU, whereas the companion found in \citet{inr03} is nearly ten times as large, 9640~AU.  This system appears to be hierarchical and is a stable configuration according to the criteria of \citet{ek95}.  The primary spectroscopic binary in this system has a metallicity measurement of $[Fe/H] = -0.31$ relative to the Sun, along with an age estimate of $3.4^{+0.3}_{-0.4}$~Gyr \citep{nord04}.}

{\noindent {\it HIP 83608C:} The companion to HIP 83608 was noted as a visual companion in the Catalogue of the Components of Double and Multiple Systems (CCDM2) \citep{ccdm2}.  Using our SpeX spectrum of this object, we found that the spectral type is definitely that of a M dwarf but that it is not a good fit to a single spectral type (see Figure \ref{fig:hip83608}).  The bluest part of the spectrum is better fit with a M4 dwarf, whereas the bulk of the spectrum more closely resembles a M1 dwarf.  Using the $M_J$-SpT relationship from \citet{haw02}, we estimated the expected spectral type, given an absolute J magnitude of 6.9$\pm$0.04, as M1.5-M2.  This agreed well with the overall M1 spectral type estimate.  However, there was still the discrepancy in the blue portion of the spectrum.  

We eliminated the possibility of this object being a background, low-metallicity subdwarf by placing it on a reduced proper motion diagram.  We used the prescription laid out in \citet{ls05} for $H_V$ versus $V-J$ where $$H_V = V + 5*log_{10}({\mu}({\arcsec}/yr)) + 5.$$  HIP 83608C has a V magnitude in the CCDM2 of 13.8; so, with a V-J color of 4.78 and a total proper motion of $0{\farcs}1164$/yr$\pm0{\farcs}043$/yr, we got a $H_V$ of 14.1.  When this was placed on Figure 30 of \citet{ls05}, it lies just above the galactic disk dwarf sequence by about a magnitude.  Thus, it is not a chance background subdwarf, and, given its slight over-brightness, it could be a binary itself.  If a binary, this system would likely be composed of an early M dwarf, given the majority of its spectrum, and a later type M dwarf as the projected absolute J magnitude matches well with the estimated spectral type.  Note that our Flamingos data do not show any elongation of the psf.  More data, particularly an optical spectrum, are required to determine the full nature of this companion.  The primary spectroscopic binary in this system has a metallicity of $[Fe/H] = -0.01$ relative to the Sun along with an age estimate of $2.2^{+0.3}_{-0.1}$~Gyr \citep{nord04}.

\noindent {{\it HIP 101769C:} The CPM object in the field of HIP 101769 was noted as a visual companion in CCDM2 and is known as CCDM J20375+1436C.  The follow-up SpeX spectrum, displayed in Figure \ref{fig:hip101769}, shows that this object is a G dwarf and is more likely to be a chance background alignment than a physical companion.  The probability of a given object being a background interloper is discussed further in Section \ref{sec:disc}.}

\subsubsection{New Tertiary Candidate Companions}

{\noindent {\it HIP 72603C:} The companion candidate to HIP 72603 has near-IR colors and absolute magnitudes consistent with an early-T dwarf.  However, it is also clearly detected in the DPOSS BRI plates.  Its SpeX prism spectrum most closely matches an early M dwarf standard (Figure \ref{fig:hip72603}).  We also performed the same reduced proper motion analysis as for HIP 83608C.  We found that this object has a $H_V$ = 16.8 and a $V-J$ = 3.1.  When placed on Figure 30 of \citet{ls05}, it falls in the subdwarf regime.  Thus, it is not a physical companion but a chance background alignment.  }

{\noindent {\it HIP 86722C:}  The new tertiary companion candidate to HIP 86722 has near-IR absolute magnitudes and colors consistent with a mid-M dwarf at the same distance as its putative primary.  It is also detected in DPOSS.  The candidate displays consistent CPM measurements over 50 years, including data from both POSS-I (1950s) and POSS-II (early 1990s).  We obtained a SpeX prism spectrum and determined its spectral type to be M4.5 (Figure \ref{fig:hip86722}).  Thus, this object appears to be a genuine physical companion.  The primary spectroscopic binary in this system has a metallicity measurement of $[Fe/H] = -0.39$ relative to the Sun \citep{nord04}.}

{\noindent {\it HIP 97944C:}  The new tertiary companion candidate to HIP 97944 has near-IR absolute magnitudes and colors consistent with a mid-M dwarf at the same distance as its putative primary.  As with HIP 86722C, this candidate is also detected in DPOSS, in both POSS-I and POSS-II, and displays consistent CPM over a baseline of 50 years.  Optical I-band photometry ($m_I = 11.94\pm0.03$~mag) from DENIS \citep{denis} and the resultant optical-NIR colors are also consistent with a mid-M dwarf.  We obtained a SpeX prism spectrum and determined its spectral type to be M5 (Figure \ref{fig:hip97944}).  Thus, we find this object to be a genuine physical companion.  The primary spectroscopic binary in this system has a metallicity measurement of $[Fe/H] = -0.11$ relative to the Sun \citep{nord04}.}

\section{Discussion}
\label{sec:disc}

\subsection{Sensitivity}
\label{sec:sens}

We made preliminary estimates of our survey sensitivity, both in separation from the bright spectroscopic binary primary and in magnitude.  This analysis was performed by creating a psf star for each field.  This psf star was then randomly placed at ten locations within the field, each with a random magnitude.  The {\it daophot} IRAF package was used to create the psf star and place the random, `fake' stars.  The resultant images were searched for point sources using identical {\it starfind} parameters from the initial search for candidates, as described in Section \ref{sec:cpmtech}.  The insertion of ten fake sources at a time was repeated 1000 times for each field for a total of 10,000 fake sources.  

The magnitude limit was then determined as a function of separation from the central binary by comparing the number of fake sources inserted to the number recovered by our search parameters.  The left-hand panel of Figure \ref{fig:sens} displays the sensitivity curve generated for the field of HIP 39064 at 50\% and 90\% completeness levels.  For this field, we see that the inner 10-15 arcseconds is mostly lost, due to the bright primary.  However, outside of $\sim$15 arcseconds, our sensitivity is fairly uniform, with an average 50\% completeness of J = 18.3 mag and a 90\% completeness of J = 17.6 mag.  In comparison to the field of HIP 39064, which corresponds to the typical brightness of our central binaries (V = 7.70 mag), the sensitivity curve of HIP 44248 (V = 3.97 mag) is displayed in the right hand panel of Figure \ref{fig:sens}.  This is one of the brightest objects in our sample.  The average 50\% completeness is 18.1 mag and the average 90\% completeness is 16.7 mag.  However, as expected for a brighter central object, the radius at which a uniform sensitivity is reached is much larger, $30-40$ arcseconds.  

It should also be noted that the effects of the bright central binaries are not limited to the central 10-15 arcseconds.  Among the brighter primaries, we noted a ringing effect in our images.  These looked like ripples in a pond, centered on the primary.  This ringing also caused wide swings in sensitivity across the FOV.  This effect can be seen in the sensitivity curve of HIP 44248 in the right-hand panel of Figure \ref{fig:sens}.  It is particularly noticeable between 20 and 50 arcsecond separations where the 90\% completeness limit jumps by several magnitudes a couple of times.  While we are not completely certain what the cause of the `ringing' is, it is likely due to scattered light within the camera from the extremely bright central objects in our fields.

Figure \ref{fig:senshist} displays a histogram of the 50\% and 90\% completeness limits for all fields observed in this work.  The median 50\% limit is 19.1 mag and the median 90\% limit is 18.2 mag.  From these data, we can see that we did not achieve our expected sensitivity in many fields (J band limit of 20 mag), but can consistently recover objects one to two magnitudes brighter.  These sensitivity curves as a function of radial separation from the central binary will be used in a future statistical analysis of the sample.

\subsection{Overall Wide Companion Rate}

Our primary goal in this initial study was to test whether known spectroscopic binaries have an enhanced, wide tertiary rate, compared to that of other stars.  Table \ref{tab:sptcf} lists the frequency of wide tertiary companions in our sample broken down by spectral type of the primary member of the spectroscopic binary system.  With the exception of the G and B stars, of which we only observed 4, the typical wide tertiary fraction is $\sim20-25\%$.  The overall rate, which we determined by assuming that these systems are drawn from a binomial distribution, is $19.5^{+5.2}_{-3.7}\%$.  This is a preliminary rate, as we do not account for incompleteness, selection effects, etc.  \citet{tok06} probes the same overall spectral type primaries as we do, but also to smaller separations.  In order to compare consistent samples we only select companions found in \citet{tok06} that we could have found in our sample.  Since we probe the same primaries we only selected companions in the same separation range ($>10\arcsec$), or 27 companions.  This yields a tertiary fraction of $16.8\pm2.9\%$ from a sample of 165 spectroscopic binaries, which is comparable to our findings.

Our wide companion fraction can also be compared to that of apparently single stars for a similar range of primary spectral type.  We also compared our companion rate to that of apparently single stars.  For example, \citet{mz04} found a wide companion rate of $\sim$11\%-12\%, which is still lower than either our own results or those of \citet{tok06}.  Thus, both this work and that of \citet{tok06} are consistent in finding an enhanced tertiary companion fraction to spectroscopic binaries with respect to `single' stars.  This trend continues into the M dwarf regime as well. \citet{law10} surveyed 36 known wide M dwarf binary systems to look for close companions to either members.  They find that $45^{+18}_{-16}\%$ of their wide binaries are actually high-order multiple systems.  This agrees with the simulation predictions of \citet{sd03}, \citet{dd04}, and \citet{umb05}, as well as with our results.  \citet{mz04} survey nearby solar type stars (FGK) over a wide range of separations and find a substellar companion rate of $\sim1\%$.  This is similar, within the uncertainties, to our observed substellar companion rate of $3.8\pm2.2\%$.  \citet{mh09} perform an exhaustive survey of 266 FGK stars with both AO imaging and wide field studies and find a substellar companion rate of $3.2^{+3.1}_{-2.7}\%$ for separations of up to $\sim$1600 AU.  This rate is statistically identical to the rate we found.  

The substellar companion fraction we report is most likely a lower limit, as we should find more faint companions when we obtain our second epoch deep imaging.  Those data will have comparable sensitivities to our first epoch ($J_{lim}{\sim}18.2$ at 90\% completeness for most fields outside of 10'' to 20'' separations) and will be sensitive to very faint T and later dwarfs.  This is because we rely on 2MASS for our first epoch astrometry.  2MASS is fairly complete to distances of 25 - 30 pc for L dwarfs \citep{kc07}.  However, T dwarfs, which all have absolute J magnitudes of $\sim 14 - 16.5$, are only detectable to distances of 15 - 20 pc, given the J-band 2MASS limit of $\sim$16.  So, at this time, it is difficult to tell if the substellar companion rate is enhanced.

Finally we examined possible correlation between the mass of the primaries and the mass of the secondaries by using spectral type as a proxy for mass.  There was no obvious correlation, although the majority of the tertiary companions were of the M spectral type for all primary spectral types.  This was not true for G type primaries, whose only confirmed companions were L dwarfs.  In any case, the large majority of wide tertiary companions noted in this work were of a considerably lower mass than their primaries, which fits with the predictions of \citet{sd03} and \citet{dd04}.

\subsection{Background Interlopers}
\label{sec:bl}

We found two background interlopers in our CPM sample, one distant G dwarf (HIP 101769C) and one background M subdwarf (HIP 72603C).  HIP 101769 has a proper motion near the limit of our minimum motion of $0\farcs1$/yr, while HIP 72603 has a proper motion much higher (${\sim}0\farcs2$/yr).  It has been found that the number density of moving objects rises as the inverse cube of the magnitude of the proper motion \citep{ls05}.  Thus, it is not surprising that the interlopers we found lie at the lower end of our motion spectrum.  

Further work by \citet{lb07} provides a quantitative mechanism for determining the likelihood that a given CPM candidate is a chance alignment of a background object when the overall motion of the object is $\ge 0{\farcs}15/yr$.  This analysis was based on the fact that the number of objects at a given proper motion increases with smaller motions and that the chance of a random alignment increases with greater angular separation.  They derived the following formula to quantify these correlations:  $${\Delta}X = [(\mu/0.15)^{-3.8}\Delta\theta\Delta\mu]^{\frac{1}{2}}$$  where $\mu$ is the magnitude of the proper motion of the primary in arcseconds per year, $\Delta\theta$ is the difference in angular position between the primary and the candidate CPM companion in arcseconds, and $\Delta\mu$ is the difference in proper motion in arcseconds per year.  The quantity ${\Delta}X$ measures the likelihood that a given object is genuine.  \citet{lb07} found that, when the value of ${\Delta}X$ is around 1, there is a 50\% chance of the candidate companion being a chance alignment.  This increases to well above 90\% for values of ${\Delta}X > 1.2$.  

The value of ${\Delta}X$ for the two interlopers we find in our sample are 1.2 for HIP 101769C and 3.9 for HIP 72603C.  Since the motion of HIP 101769C is below the limit of the analysis in \citet{lb07}, it is not clear how effectively this formula can be applied.  All of our other candidates have values under 1.

\subsection{Spectroscopic Binary Periods}

The comparison between the target sample of this work and that of \citet{tok06} produced some interesting results, particularly when comparing the orbital periods of the target spectroscopic binaries.  Tokovinin's program set out with the same goal as ours: examination of the tertiary fraction of spectroscopic binaries as a means of testing binary star formation simulations.  They wanted to maximize the chances of detecting tertiaries; so, they selected only spectroscopic binaries that only have periods of less than 30 days.  The simulation predictions of \citet{sd03} and \citet{dd04} argue that the Kozai mechanism can tighten these systems by transferring angular momentum from the tight system to a wide third member.  Thus, the tighter systems should, preferentially, have a higher tertiary companion rate than wider spectroscopic binaries.  In the \citet{tok06} study they do find that tighter spectroscopic binaries tend to have a higher fraction of tertiaries and that the tertiary fraction rises to nearly 100\% for spectroscopic binaries with periods of less than a day.

Our selected sample is volume and minimum proper motion limited, while the period of the spectroscopic binary is unconstrained.  Table \ref{tab:percf} lists companion fraction as a function of the period of the spectroscopic binaries in the observed sample, while Figure \ref{fig:percf} displays a histogram of the data in Table \ref{tab:percf}.  It is broken down in log period bins centered on the values listed with $\pm0.5~dex$ widths.  We calculated the fractions, assuming that the data are binomially distributed (Table \ref{tab:percf}).  Our number of primaries as a function of binary period is fairly flat from 1 day to 10,000 days.  The sample from \citet{tok06} does not examine spectroscopic binaries with periods longer than 30 days, which corresponds to the last three bins of Table \ref{tab:percf} and Figure \ref{fig:percf}.  The tertiary companion rate, however, peaks at the smallest period bin for which we have significant data, $50\% \pm 13.3\%$ for spectroscopic binaries with periods between 1 and 10 days.  The companion fraction then decreases for spectroscopic binaries with periods between 10 and 100 days, but then increases again to between 15\% and 30\%.  Note that the most significant of these variations from our baseline 19.5\% companion rate is the 50\% rate at small periods, which is a ${\sim}2\sigma$ event.  Thus, we can marginally confirm the result of \citet{tok06}, which found that the tertiary companion rate drops by about a factor of two from very close binaries (periods less than 7 days) to wide binaries (periods between 7 and 30 days).  A larger sample size is needed to provide a more robust measurement of the variation of the companion fraction as a function of spectroscopic binary period.  \citet{tok06} also find that there is no correlation between the period of the binary and the separation of the tertiary, which we find as well (see Figure \ref{fig:pvs}).  However, our result of an increase in companion fraction for spectroscopic binaries with periods longer than 100 days is quite different from that of \citet{tok06}.  

We found a significant fraction of wide spectroscopic binaries have tertiary companions.  This demonstrates that enhanced wide companion rates also apply to wider spectroscopic binaries (Table \ref{tab:percf}).  However, this result is preliminary, as there is a large area of separation space that this work has not yet explored, particularly tertiary companions with separations less than 10\arcsec.  It should be noted that \citet{tok06} found nearly half of their tertiary companions in this separation range.

\section{Summary}

We surveyed a volume-limited sample of 77 spectroscopic binaries for CPM tertiary companions.  We found or confirmed a total of 4 tertiary companions, two of which are brand new to this work.  When combined with the 11 previously known companions, this yields a total of 15 companions in 13 triple systems and 1 quadruple (HIP 41211), for an observed wide companion rate of $19.5\%^{+5.3\%}_{-3.7\%}$.  Note that these numbers are valid for separations $>10-20$ arcseconds or hundreds of AU.  Also, our magnitude limit is up to $3-4$ magnitudes deeper than 2MASS; thus, we are currently sensitive only to sources also detected in 2MASS.

Initial comparisons with other observed tertiary or wide binary rates yield contradictory results.  First, when we compared these data to the only other similar survey, \citet{tok06}, we found a similar tertiary companion fraction.  When we compared these fractions to `single' FGK stars, of which the bulk of our sample binaries are composed, we found an enhanced fraction of wide companions ($\sim$11\%-12\% from \citet{mz04}, versus 19.5\% from our work).  However, when comparing the substellar tertiary rate of this sample, $3.8\pm2.2\%$, to that of similar surveys \citep{mz04,mh09}, 3.2\%, we found little difference, though it should be noted that all cases deal with small-number statistics.  These results are preliminary and will require additional analysis to confirm.  In particular, a second epoch of deep imaging will be needed to search for fainter tertiary companions, which include objects in the Y dwarf regime \citep{davy08,cush11}.  

{\it Acknowledgments:}  P. A. was supported by grant NAG5-11627, awarded to Kevin Luhman from the NASA Long-Term Space Astrophysics program.  This research has made use of the NASA/IPAC Infrared Science Archive, which is operated by the Jet Propulsion Laboratory, California Institute of Technology, under contract with the National Aeronautics and Space Administration.  The work here also made use of the Second Palomar Observatory Sky Survey (POSS-II).  POSS-II was made by the California Institute of Technology with funds from the National Science Foundation, the National Geographic Society, the Sloan Foundation, the Samuel Oschin Foundation, and the Eastman Kodak Corporation.  This publication makes use of data products from the Two Micron All Sky Survey, which is a joint project of the University of Massachusetts and the Infrared Processing and Analysis Center/California Institute of Technology, funded by the National Aeronautics and Space Administration and the National Science Foundation.  This research has made use of the SIMBAD database, operated at CDS, Strasbourg, France.  P.A.\ would also like to thank the many individuals who obtained the additional SpeX NIR prism spectra presented here: Kelle Cruz, Dagny Looper, and Kevin Luhman.  P.A.\ acknowledges several productive conversations that aided the work presented here, including Sebastien L{\'e}pine, Adam Kraus, and Kevin Luhman.  The authors would finally like to thank the observatory staffs at KPNO and CTIO for their help in obtaining the data used in this paper, particularly Dick Joyce and Nicole van der Bliek.

\clearpage

\begin{figure}
\begin{center}
\includegraphics[scale=0.40,angle=-90]{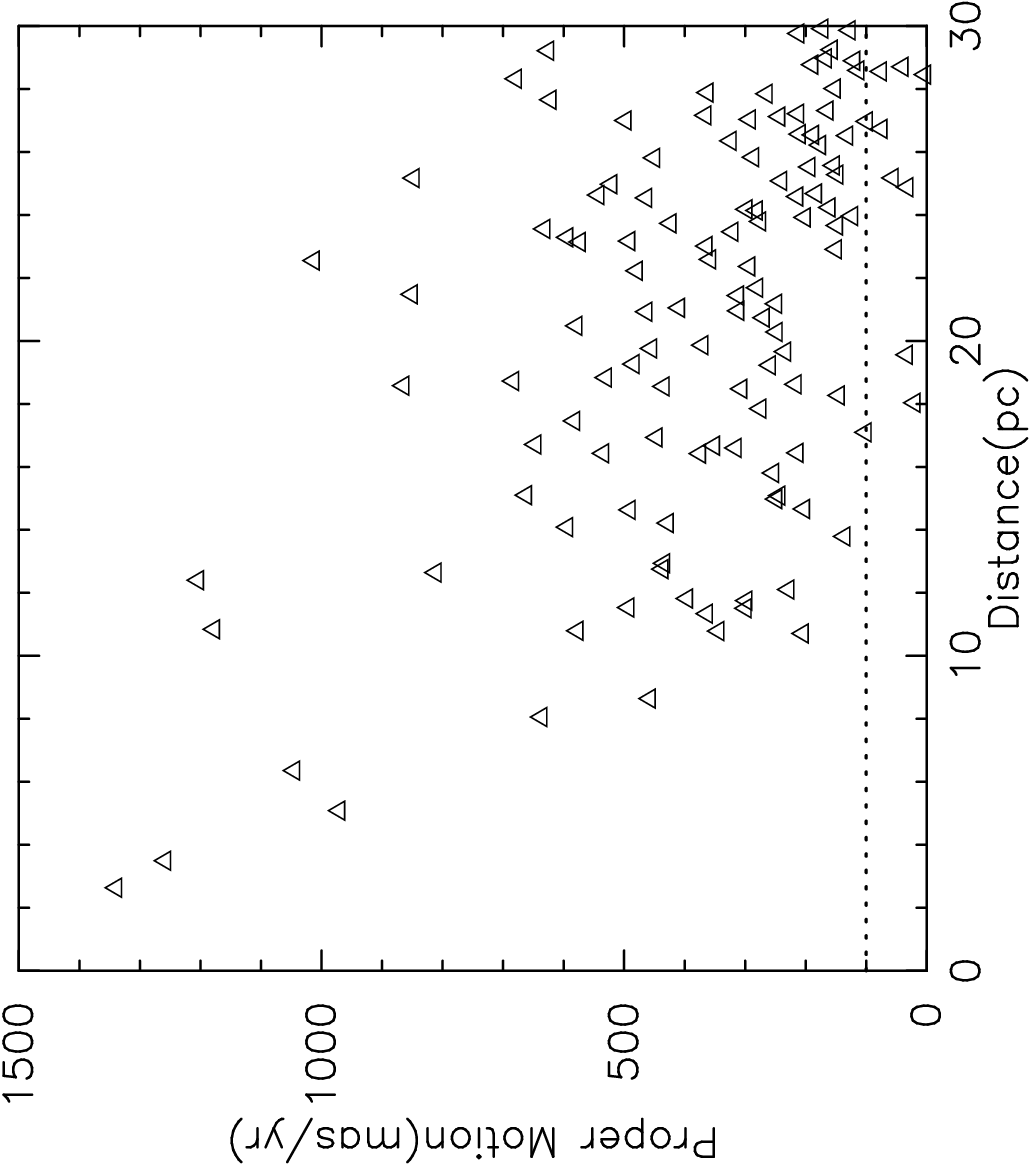}
\includegraphics[scale=0.40,angle=-90]{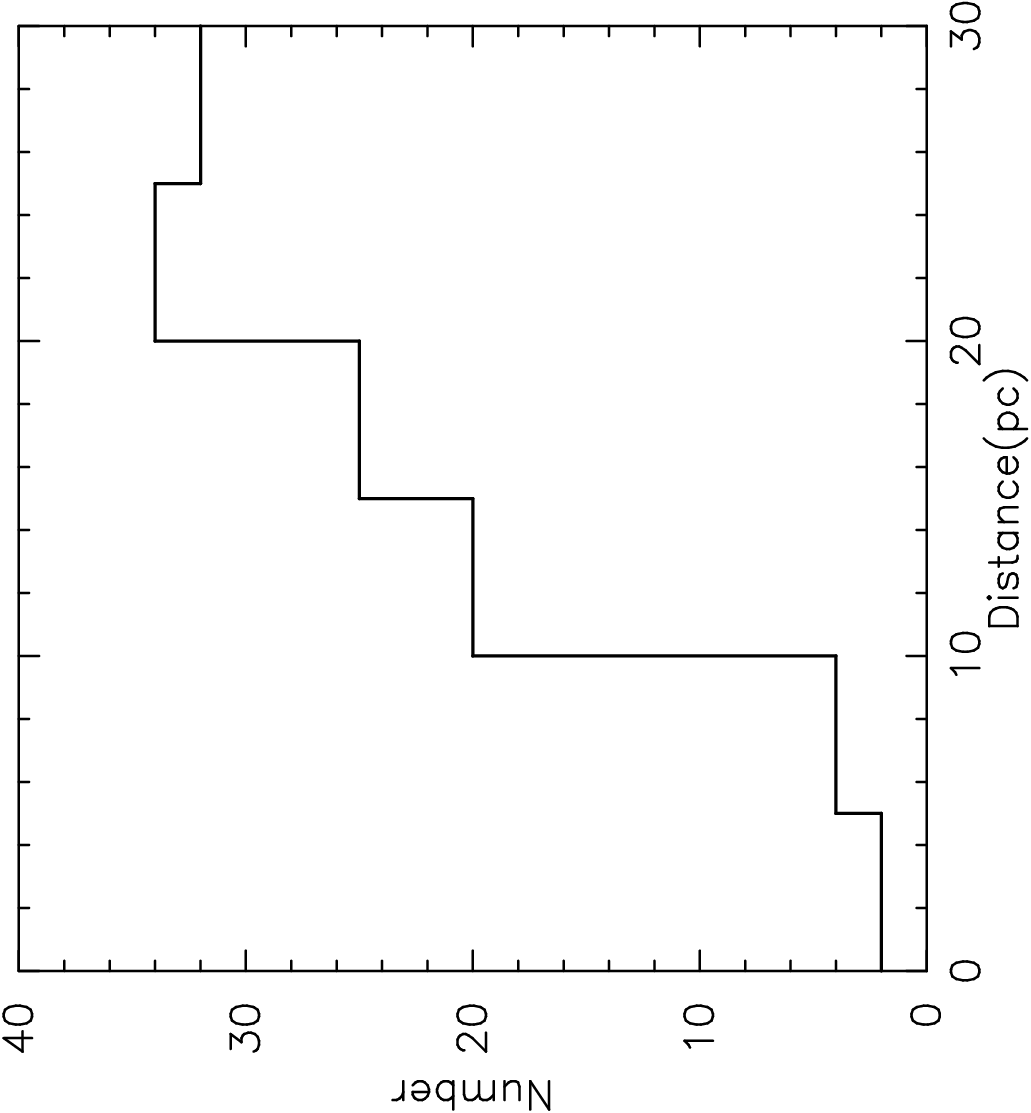}
\end{center}
\caption{Left: Measured distances versus proper motions for all Hipparcos spectroscopic binaries within 30~pc of the Sun.  Open triangles represent each object in the sample.  The horizontal dotted line gives the minimum proper motion cut-off for inclusion in the final target list.  Right:  Cumulative histogram of the number of spectroscopic binaries as a function of distance.  Note that the number flattens at distances larger than 20~pc, indicating that this sample is volume-limited and not volume-complete.}
\label{fig:pmvd}
\end{figure}

\begin{figure}
\begin{center}
\includegraphics[scale=1.0,angle=-90]{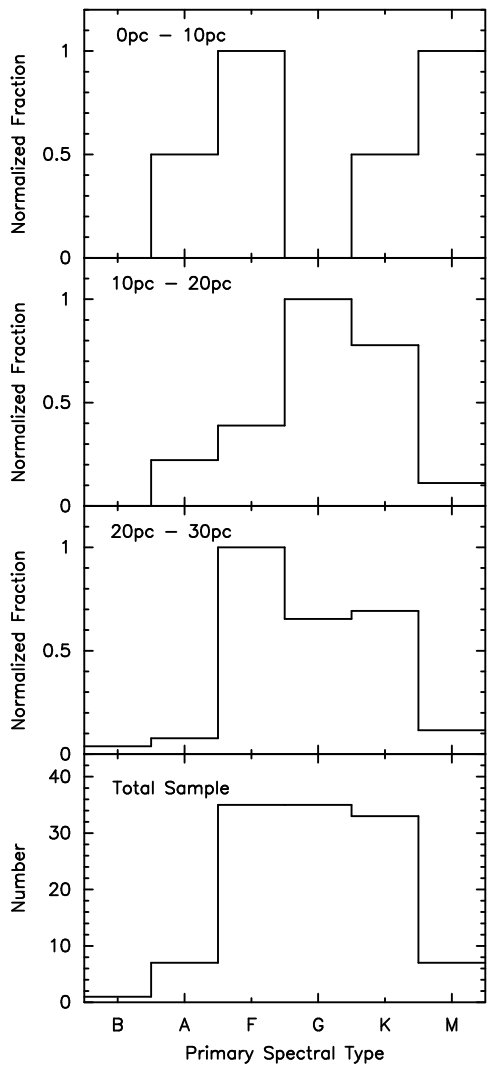}
\end{center}
\caption{Histograms of the spectroscopic binary spectral types in the sample we derived from the SB9 catalog \citep{sb9}.  The bottom panel shows our entire sample, note that we have a large number of FGK stars but very few B, A, or M stars, and no O stars.  The top three panels show histograms of the relative fraction of primary spectral types in our sample broken down into three distance bins: 0 pc - 10 pc (solid red); 10 pc - 20 pc (dashed green); 20 pc - 30 pc (dot-dashed blue).  Note that the peak of the distribution shifts to earlier spectral types with increasing distance.}
\label{fig:sptdist}
\end{figure}

\begin{figure}
\begin{center}
\includegraphics[scale=0.40,angle=-90]{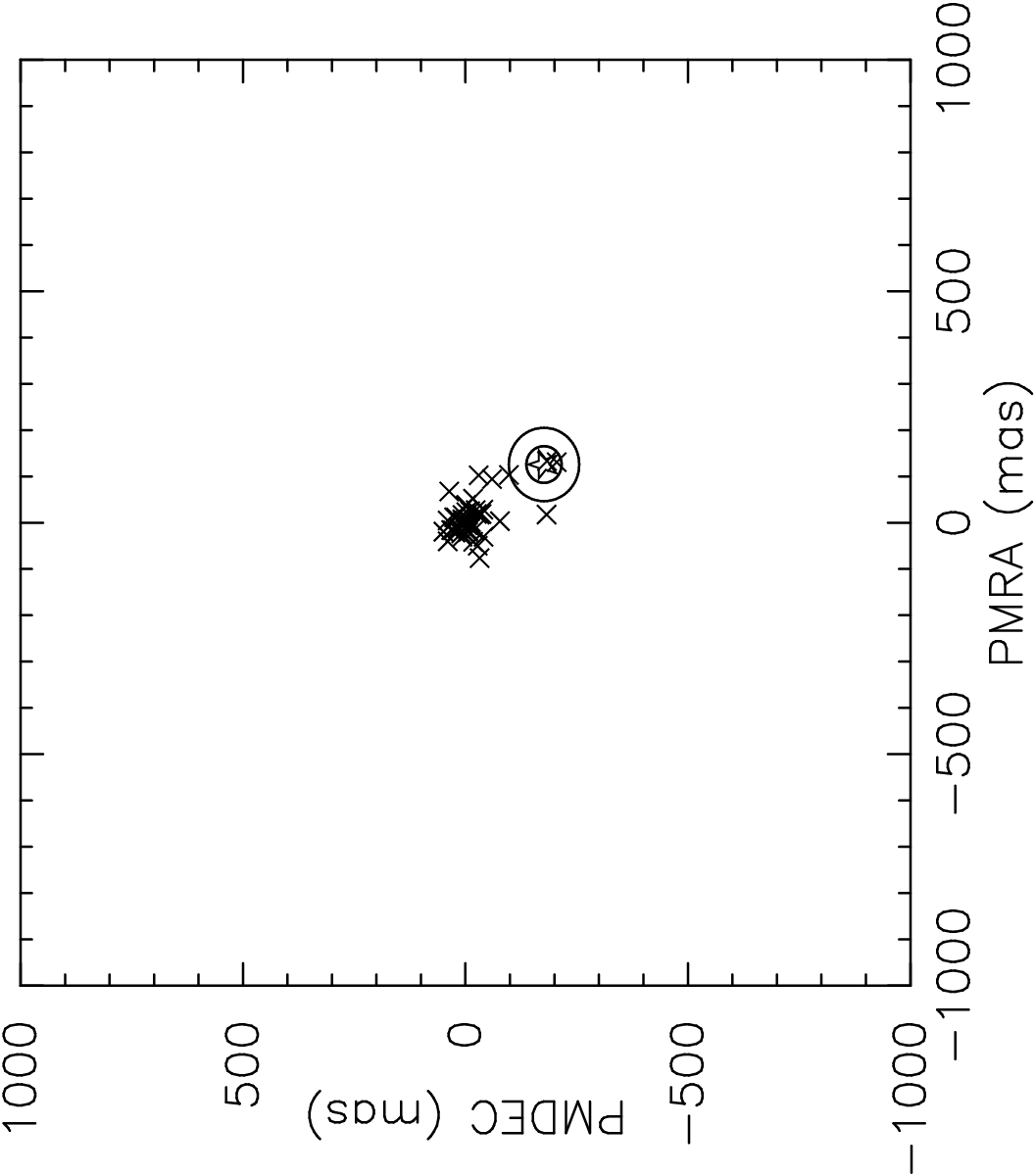}
\includegraphics[scale=0.40,angle=-90]{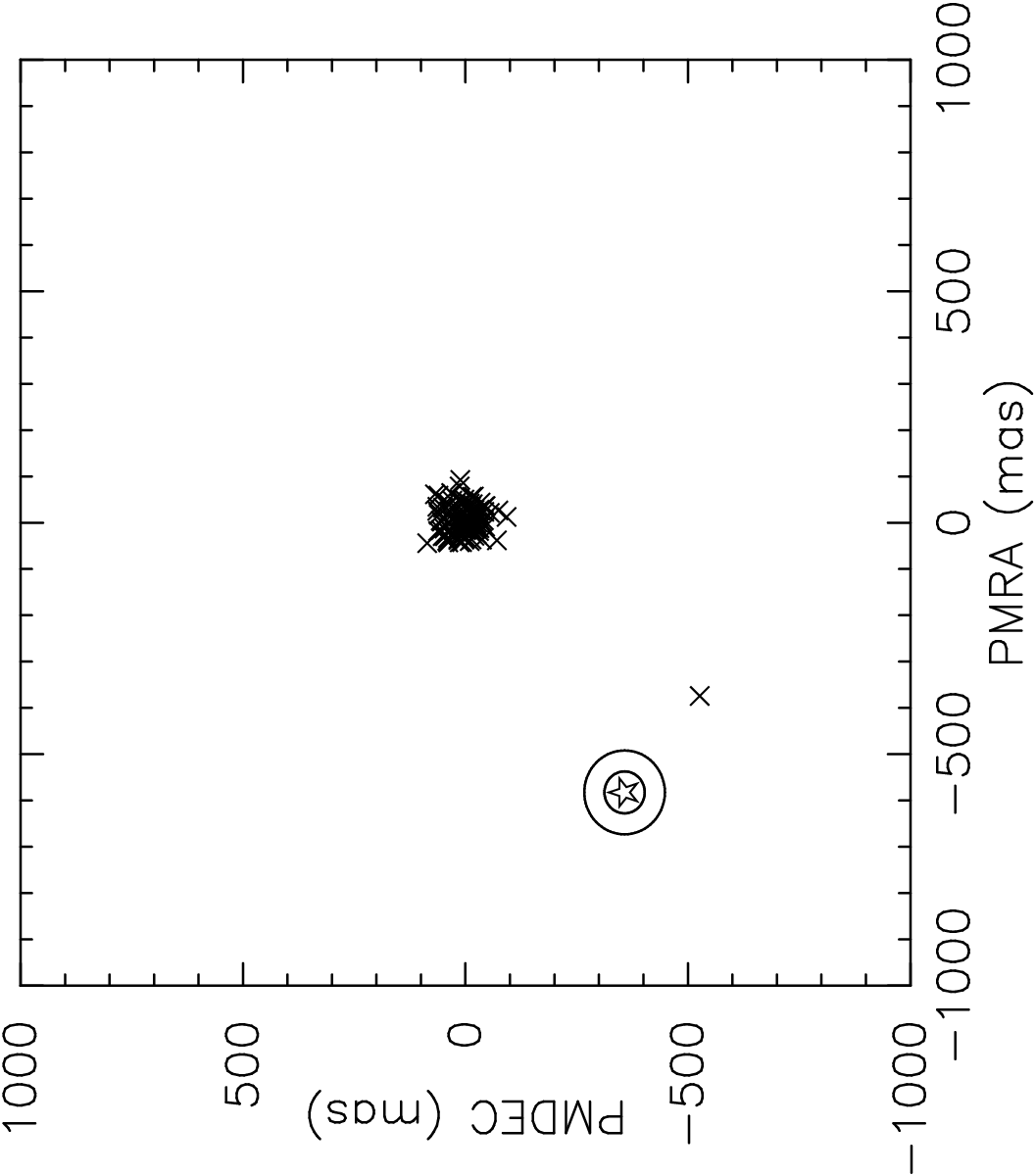}
\end{center}
\caption{Representative plots of the CPM measurements for the fields around two target spectroscopic binaries.  x's represent the measured motion of each object in the field detected in both the 2MASS PSC and in the ISPI/Flamingos data.  The five pointed star indicates the proper motion of the primary, and the two large circles are the $1 \sigma$ and $2 \sigma$ detection thresholds for candidate co-moving companions measured in mas/yr.  The left panel is that of the field around HIP 75312 and it contains a known L dwarf tertiary.  The right panel is that of HIP 105312 and is an example of a field without a companion.  The HIP 105312 field also typifies the mismatches that lead to spurious high proper motion candidates discussed in Section \ref{sec:cpmtech}.  Note the source that is not clustered around zero motion; it is a spurious detection.  However, since it is not within $1\sigma-2\sigma$ of the primary it is not counted in this case.}
\label{fig:cpm}
\end{figure}


\begin{figure}
\begin{center}
\includegraphics[scale=0.6,angle=-90]{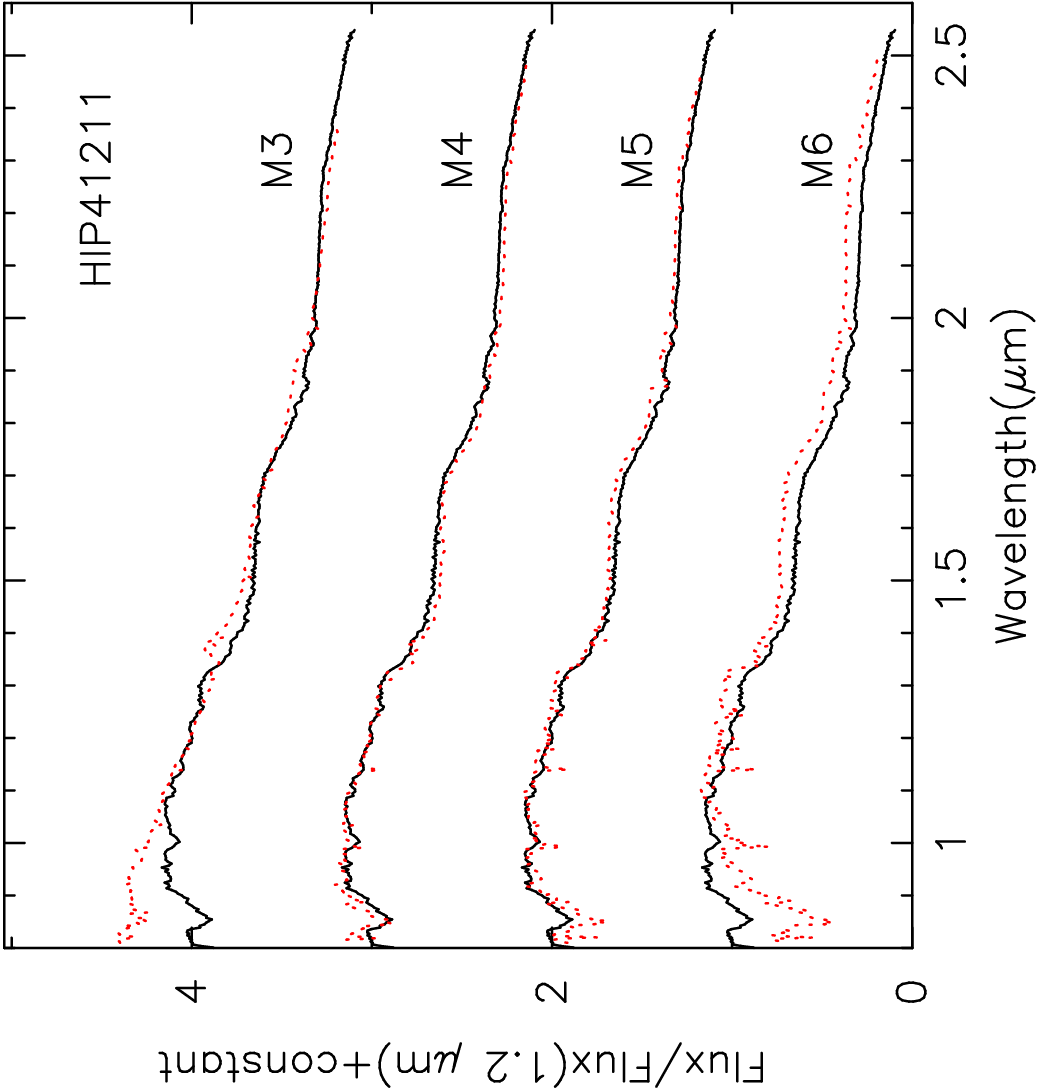}
\end{center}
\caption{SpeX prism spectrum of the candidate companion to HIP 41211 (black solid line) compared to M3-M6 dwarf spectra from the IRTF Spectral Library (red dotted line) \citep{cush05,rayn08}.  We find that the candidate is a M4.5 dwarf, based on this spectrum.  Thus, coupled with the measured magnitudes and colors (see Table \ref{tab:cand}), it is consistent with being a {\it bona fide} companion.}
\label{fig:hip41211}
\end{figure}

\begin{figure}
\begin{center}
\includegraphics[scale=0.4,angle=-90]{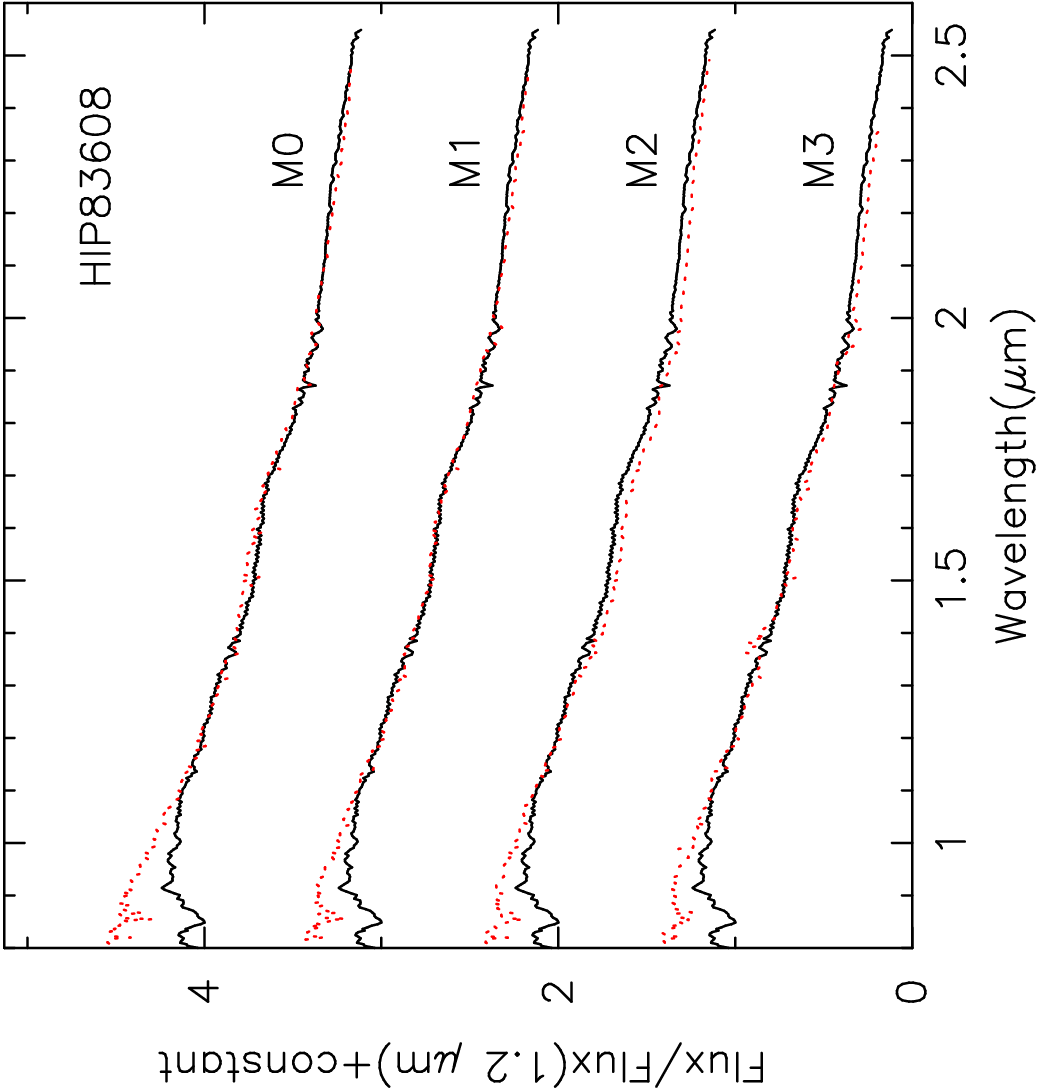}
\includegraphics[scale=0.4,angle=-90]{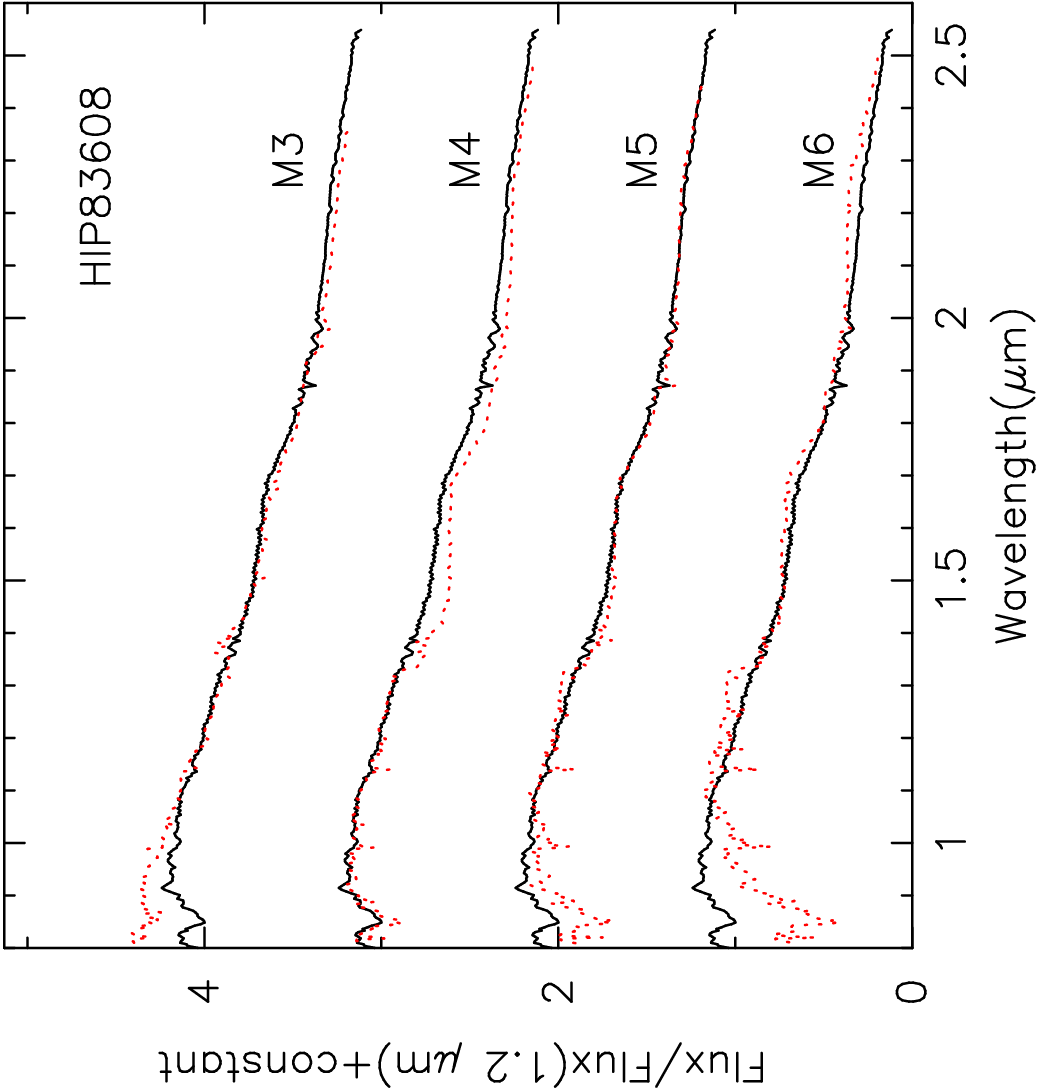}
\end{center}
\caption{SpeX prism spectrum of the candidate companion to HIP 83608 (black solid line) compared to standard spectra (red dotted line), M0-M3 in the left panel and M3-M6 in the right panel.  We obtained the standards from the IRTF Spectral library  \citep{cush05,rayn08}.  The overall morphology of the spectrum is that of a M1 dwarf in the red portion of the spectrum and a M4 dwarf in the blue.  It could be a composite of an early and a mid M dwarf and, thus, a binary itself.}
\label{fig:hip83608}
\end{figure}

\begin{figure}
\begin{center}
\includegraphics[scale=0.6,angle=-90]{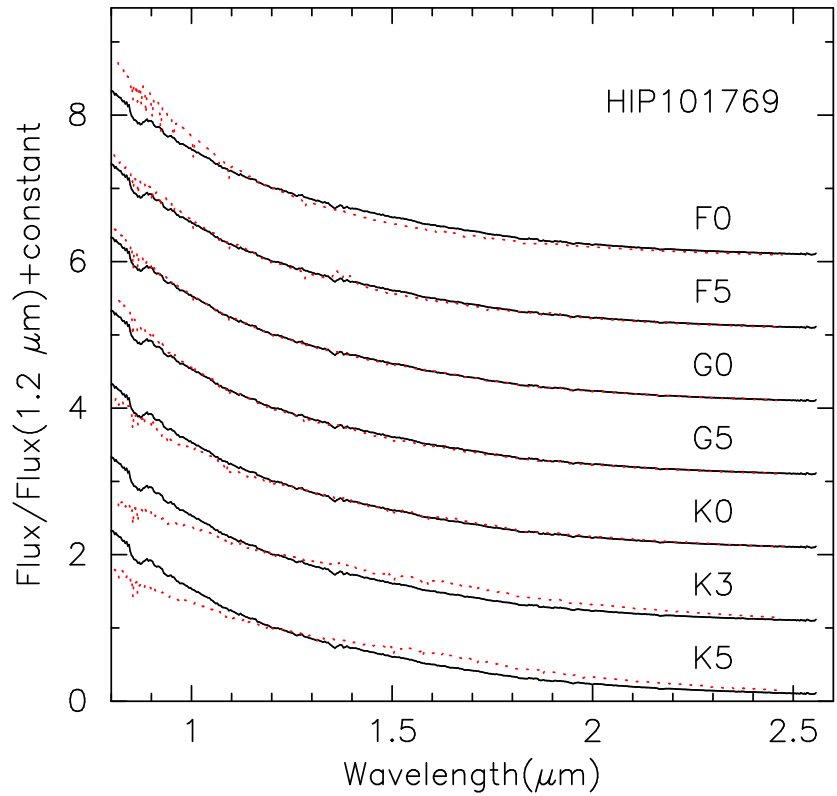}
\end{center}
\caption{SpeX prism spectrum of the candidate companion to HIP 101769 (black solid line) compared to standard stellar spectra from the IRTF Spectral Library \citep{cush05,rayn08} (red dotted lines).  This object is clearly a late-G dwarf and, thus, cannot be a {\it bona fide} companion, given its projected absolute magnitudes and colors (Table \ref{tab:cand}).}
\label{fig:hip101769}
\end{figure}

\begin{figure}
\begin{center}
\includegraphics[scale=0.6,angle=-90]{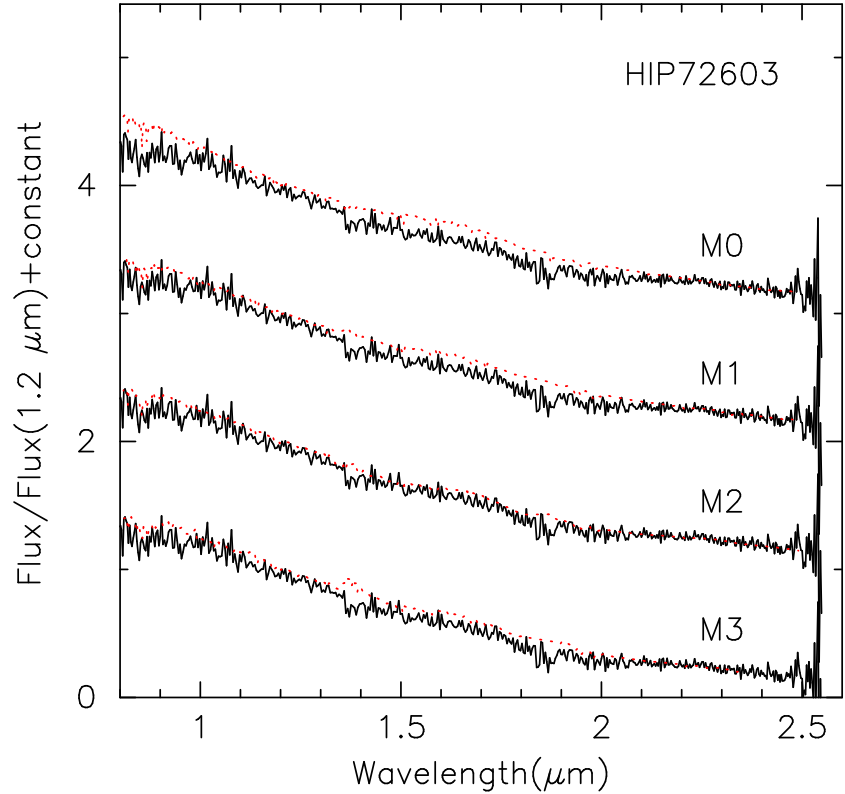}
\end{center}
\caption{SpeX prism spectrum of the candidate companion to HIP 72603 (black solid line), compared to M0-M3 dwarf spectra (red dotted lines) from the IRTF Spectral Library standards \citep{cush05,rayn08}.  The overall morphology of the spectrum is that of a M dwarf, but it is not a good fit to any particular template.  Coupled with the reduced proper motion determination (Section \ref{sec:bl}), it is likely a unrelated background subdwarf.}
\label{fig:hip72603}
\end{figure}

\begin{figure}
\begin{center}
\includegraphics[scale=0.6,angle=-90]{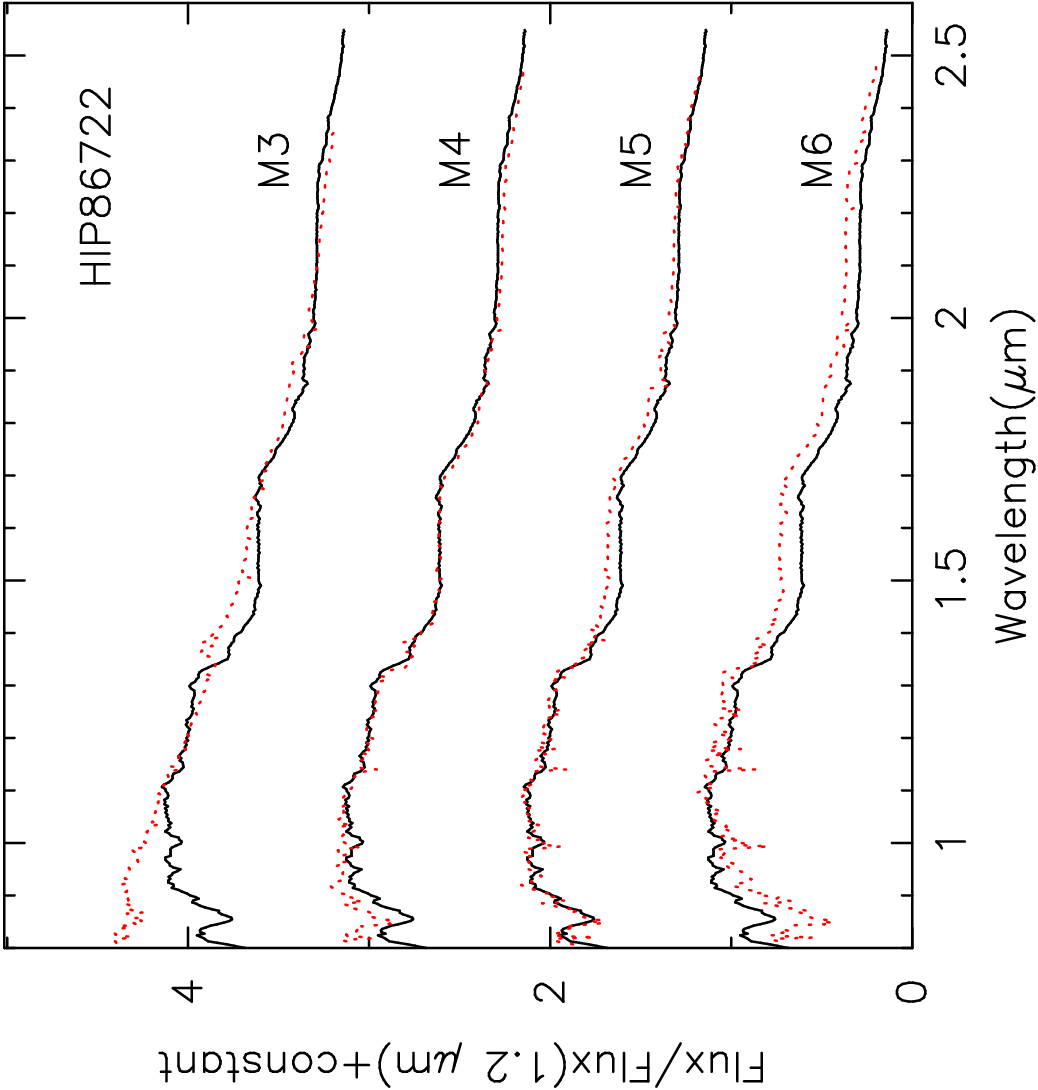}
\end{center}
\caption{SpeX prism spectrum of the candidate companion to HIP 86722 (black solid line) compared to M3-M6 dwarf spectra (red dotted lines) from the IRTF Spectral Library \citep{cush05,rayn08}.  We find that the candidate is a M4.5 dwarf based on this spectrum.  Thus, coupled with the measured magnitudes and colors (see Table \ref{tab:cand}), it is consistent with being a {\it bona fide} companion.}
\label{fig:hip86722}
\end{figure}

\begin{figure}
\begin{center}
\includegraphics[scale=0.6,angle=-90]{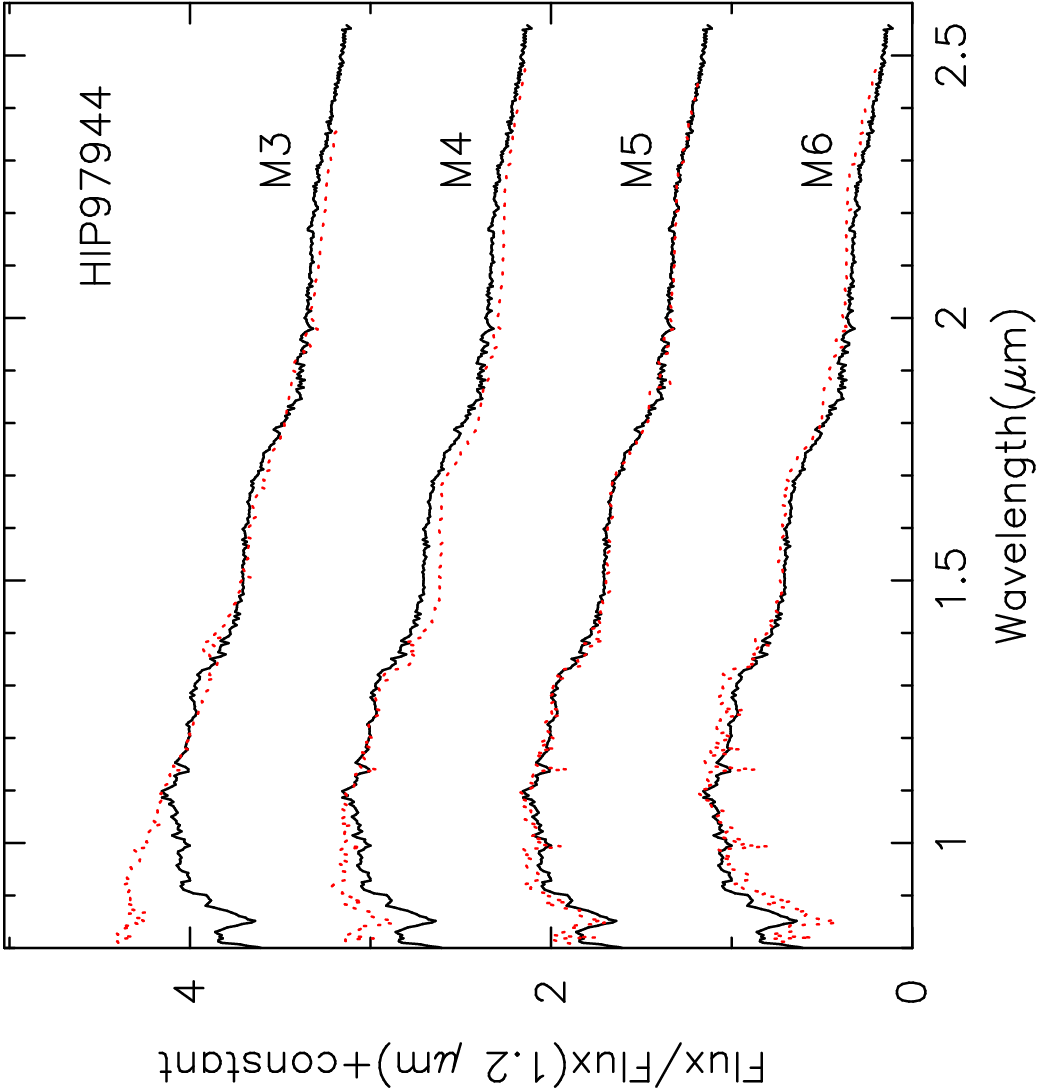}
\end{center}
\caption{SpeX prism spectrum of the candidate companion to HIP 97944 (black solid line) compared to M3-M6 dwarf spectra (red dotted lines) from the IRTF Spectral Library \citep{cush05,rayn08}.  We find that the candidate is a M5 dwarf based on this spectrum.  Thus, coupled with the measured magnitudes and colors (see Table \ref{tab:cand}), it is likely a {\it bona fide} companion.}
\label{fig:hip97944}
\end{figure}

\begin{figure}
\begin{center}
\includegraphics[scale=0.40,angle=-90]{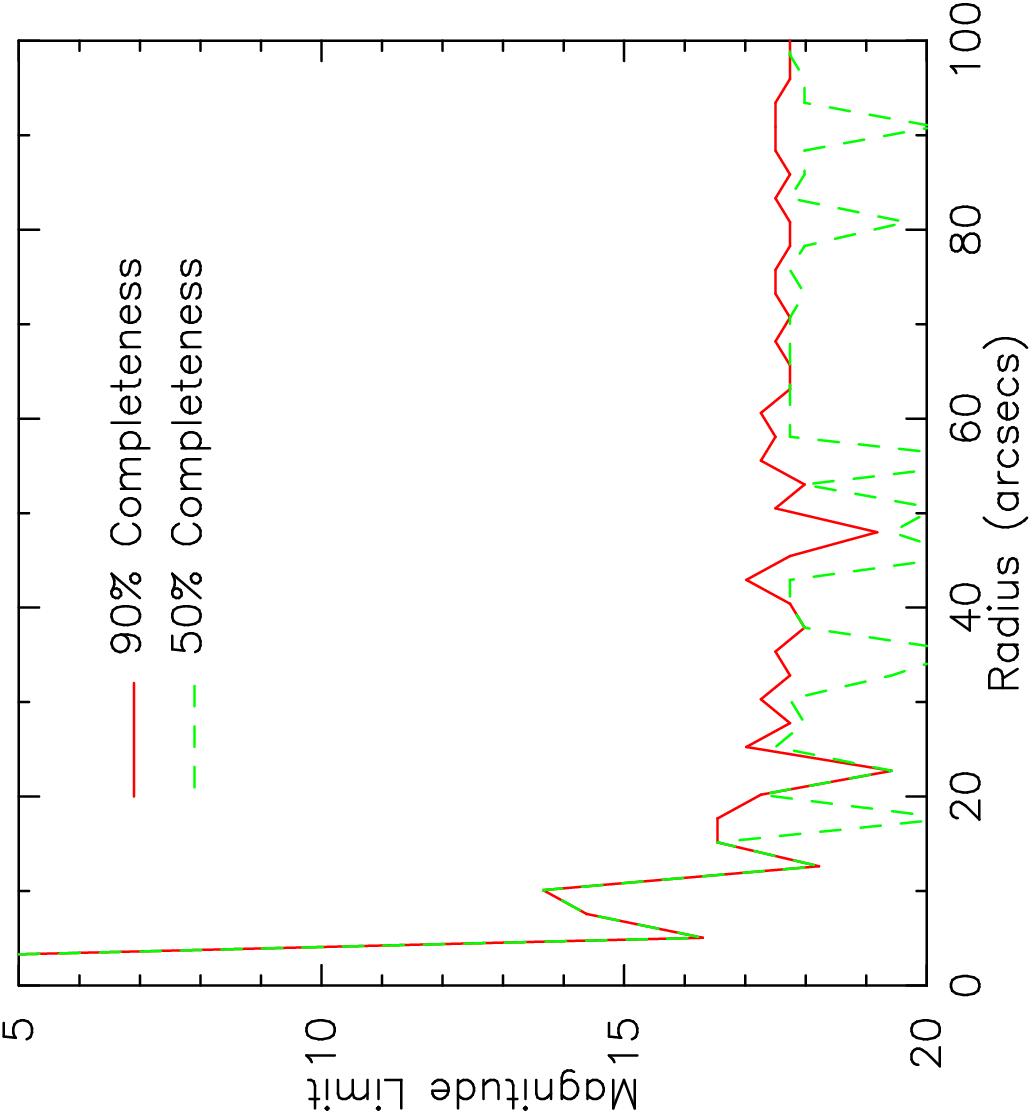}
\includegraphics[scale=0.40,angle=-90]{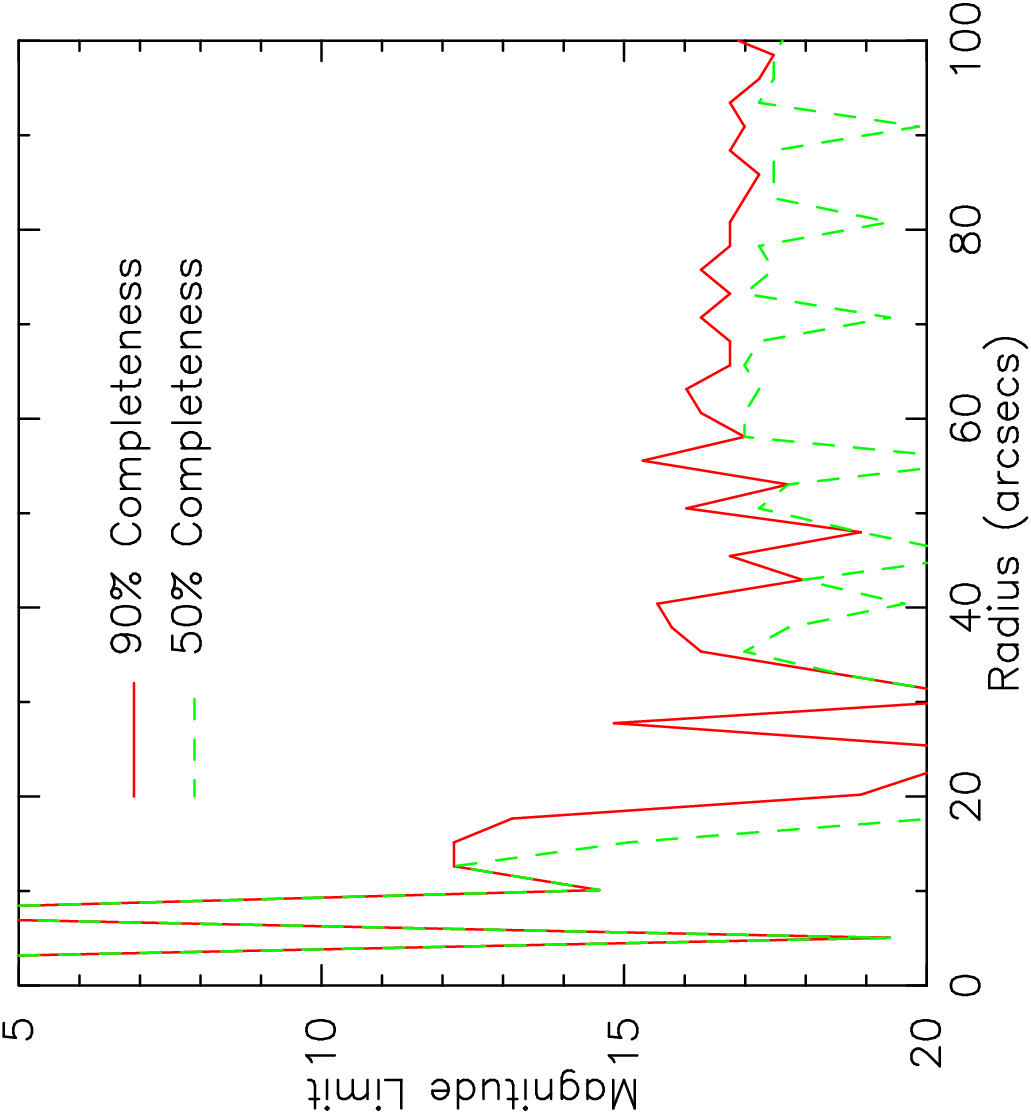}
\end{center}
\caption{J-band magnitude limits as a function of radial separation from the central spectroscopic binaries for the fields of HIP 39064 (left) and HIP 44248 (right), with 50\% completeness displayed as the solid red line and 90\% completeness as the dashed green line.  HIP 39064 (V = 7.7 mag near the median value of primary magnitudes for our survey), represents our typical sensitivity, whereas HIP 44248 is at the bright end of our sample, with V = 4.0 mag.  Note that, as expected, the sensitivity as a function of separation is degraded for the brighter star.}
\label{fig:sens}
\end{figure}

\begin{figure}
\begin{center}
\includegraphics[scale=0.6,angle=-90]{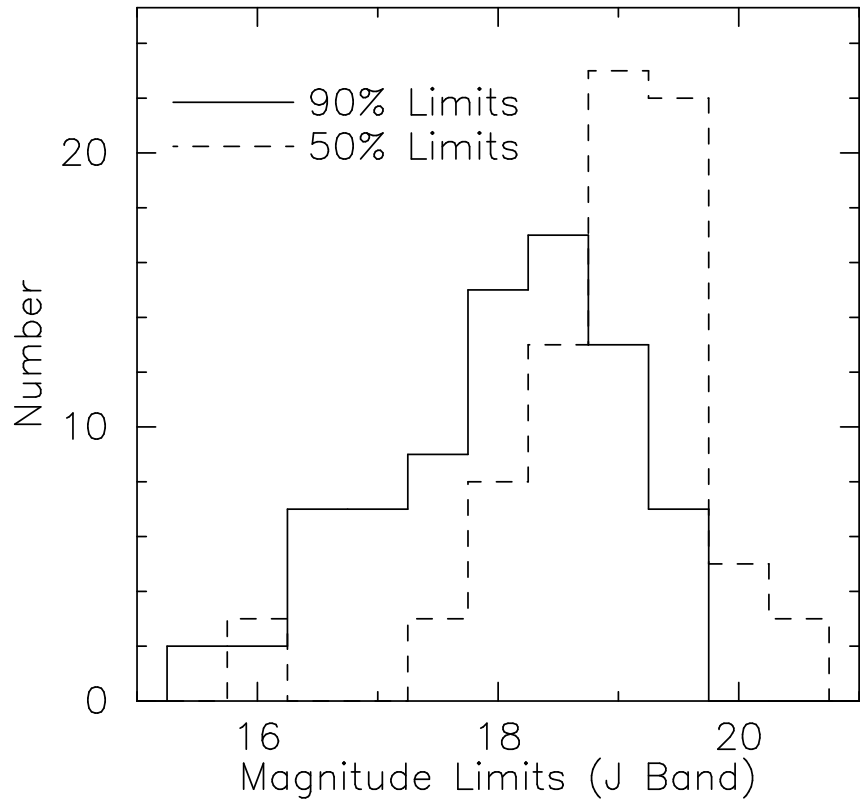}
\end{center}
\caption{Histogram of 50\% and 90\% completeness J band limiting magnitudes for all observed fields.  There is a wide range of completeness limits, due to the changes in seeing and primary brightness across all the observed fields.  The median J band completeness limit at 50\% is 19.1 mag and at 90\% is 18.2 mag.}
\label{fig:senshist}
\end{figure}

\begin{figure}
\begin{center}
\includegraphics[scale=0.6,angle=-90]{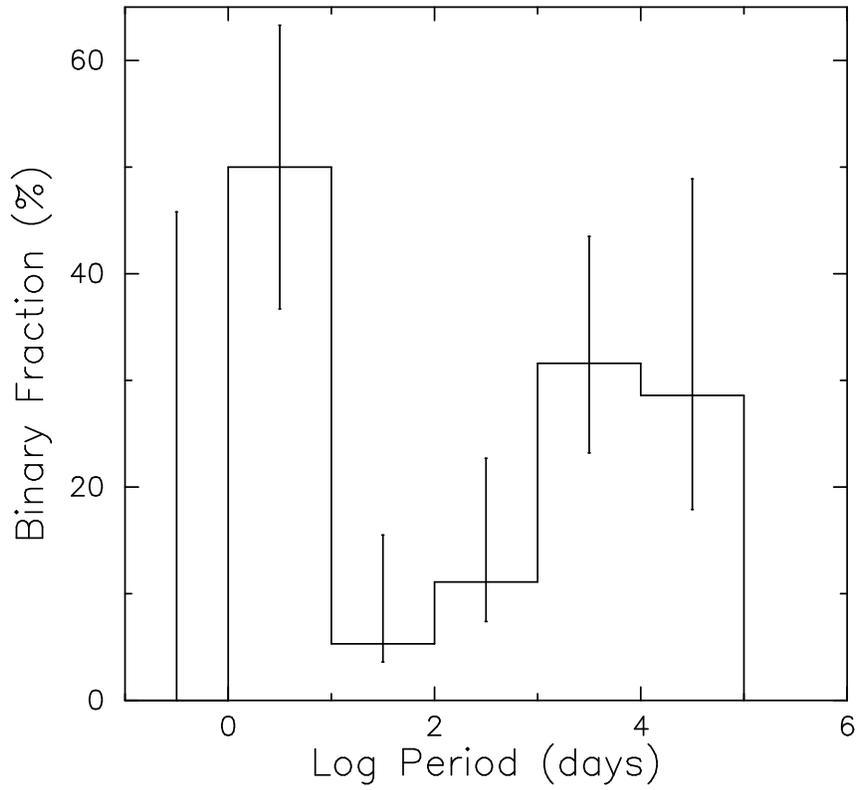}
\end{center}
\caption{Histogram of the period of the central spectroscopic binaries (in Log days) versus the calculated companion frequency.  Note the large spike in companion frequency at the smallest binary periods.  This is consistent with the predictions of \citep{sd03} and \citep{dd04}.}
\label{fig:percf}
\end{figure}

\begin{figure}
\begin{center}
\includegraphics[scale=0.6,angle=-90]{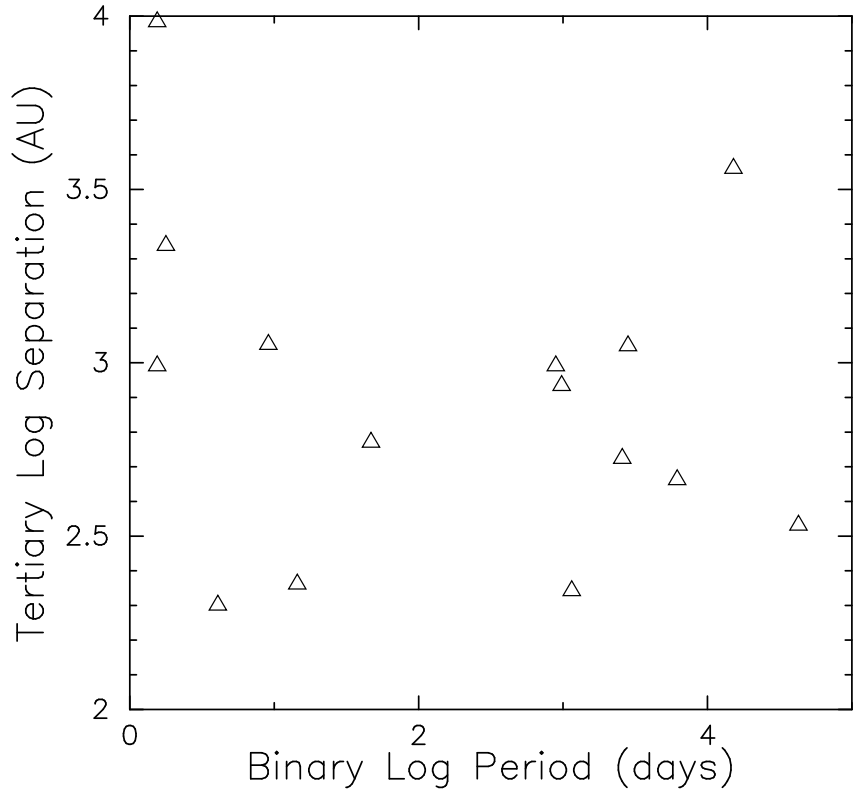}
\end{center}
\caption{Scatter plot of the period of the central spectroscopic binaries with tertiaries against the projected separation of its tertiary companion.  Note that there is little visual evidence of a correlation between them.}
\label{fig:pvs}
\end{figure}

\begin{deluxetable}{ccccccrrccc}
\tabletypesize{\footnotesize}
\rotate
\tablewidth{0pt}
\tablecaption{Spectroscopic Binary Target List}
\tablehead{
\colhead{Hipparcos} & \colhead{RA} & \colhead{DEC} & \colhead{$\pi$\tablenotemark{a}} & \colhead{$\mu_{\alpha}\tablenotemark{a}$} & \colhead{$\mu_{\delta}\tablenotemark{a}$} & \colhead{Period\tablenotemark{b}} & \colhead{V\tablenotemark{b}} & \colhead{SpT\tablenotemark{c}} & \colhead{UT Date Obs.} & \colhead{Telescope} \\
\colhead{Number} & \colhead{(h am as)} & \colhead{(d m s)} & \colhead{(mas)} & \colhead{(mas/yr)} & \colhead{(mas/yr)} & \colhead{(days)} & \colhead{(mag)} & \colhead{} & \colhead{} & \colhead{} }

\startdata

    171 & 00 02 09.65 & +27 05 04.2 & $ 80.63\pm3.03$ & $  778.59\pm2.81$ & $ -918.72\pm1.81$ &  9594.910 &  5.75 &        G2V &  &  \\
    518 & 00 06 15.81 & +58 26 12.2 & $ 49.30\pm1.05$ & $  247.36\pm0.81$ & $   17.77\pm0.70$ &    47.510 &  8.00 &        G9V &  &  \\
    677 & 00 08 23.17 & +29 05 27.0 & $ 33.60\pm0.73$ & $  135.68\pm0.63$ & $ -162.95\pm0.45$ &    96.696 &  2.17 &        A0p &  &  \\
   1349 & 00 16 53.59 & -52 39 05.7 & $ 43.45\pm1.19$ & $  314.94\pm0.72$ & $  182.50\pm0.69$ &   411.449 &  6.84 &        G2V & 23 Jan  2008 & CTIO \\
   2081 & 00 26 16.87 & -42 18 18.4 & $ 42.14\pm0.78$ & $  232.76\pm0.54$ & $ -353.64\pm0.82$ &  3848.830 &  2.40 &      K0III & 21 Jan  2008 & CTIO \\
   2762 & 00 35 14.88 & -03 35 34.2 & $ 47.51\pm1.15$ & $  407.68\pm1.31$ & $  -36.47\pm0.61$ &  2527.000 &  5.20 &        F8V &  &  \\
   3362 & 00 42 48.06 & +35 32 55.0 & $ 42.03\pm1.98$ & $  264.99\pm1.81$ & $   74.49\pm1.43$ &     2.170 & 10.38 &       dM1e &  &  \\
   3810 & 00 48 58.71 & +16 56 28.1 & $ 41.80\pm0.75$ & $   -2.33\pm0.71$ & $ -201.57\pm0.55$ &    13.825 &  5.07 &        F8V &  &  \\
   3850 & 00 49 26.76 & -23 12 44.9 & $ 53.09\pm1.02$ & $  516.74\pm1.04$ & $  119.52\pm0.72$ &  6832.000 &  7.17 &     G8/K0V & 22 Jan  2008 & CTIO \\
   6917 & 01 29 04.90 & +21 43 23.4 & $ 43.16\pm0.93$ & $  455.58\pm1.07$ & $ -185.00\pm0.84$ &    10.984 &  7.74 &        K2V &  &  \\
   7078 & 01 31 13.52 & +70 15 53.2 & $ 34.19\pm0.57$ & $  137.53\pm0.49$ & $  -76.00\pm0.50$ &   134.078 &  5.82 &        F6V &  &  \\
   7918 & 01 41 47.14 & +42 36 48.1 & $ 79.09\pm0.83$ & $  791.35\pm0.65$ & $ -180.16\pm0.47$ &  7122.000 &  4.90 &        G2V &  &  \\
   8796 & 01 53 04.90 & +29 34 45.8 & $ 50.87\pm0.82$ & $   12.02\pm0.67$ & $ -233.69\pm0.51$ &     1.767 &  3.41 &       F6IV &  &  \\
   8903 & 01 54 38.35 & +20 48 29.9 & $ 54.74\pm0.75$ & $   96.32\pm1.00$ & $ -108.80\pm0.54$ &   106.997 &  2.60 &        A5V &  &  \\
  10644 & 02 17 02.42 & +34 13 29.4 & $ 92.20\pm0.84$ & $ 1151.61\pm0.87$ & $ -246.32\pm0.77$ &    10.020 &  4.87 &        G0V &  &  \\
  10723 & 02 18 01.23 & +01 45 24.8 & $ 40.04\pm0.92$ & $  365.99\pm1.08$ & $  371.16\pm0.66$ &    93.500 &  5.56 &        F9V & 24 Jan  2008 & CTIO \\
  11349 & 02 26 01.70 & +05 46 46.0 & $ 35.86\pm1.15$ & $  352.45\pm1.15$ & $   83.14\pm1.17$ &  3600.000 &  7.95 &        G5V &  &  \\
  11964 & 02 34 22.52 & -43 47 44.3 & $ 86.87\pm0.86$ & $   57.89\pm0.82$ & $ -293.65\pm0.72$ &     1.562 &  8.70 &       K7Ve & 21 Jan  2008 & CTIO \\
  12390 & 02 39 33.73 & -11 52 17.7 & $ 36.99\pm1.76$ & $  172.01\pm1.55$ & $ -236.31\pm1.31$ &   975.900 &  4.84 &     F5IV-V & 24 Jan  2008 & CTIO \\
  12623 & 02 42 14.93 & +40 11 39.8 & $ 40.52\pm1.25$ & $  -16.84\pm0.84$ & $ -182.29\pm0.88$ &   331.000 &  4.92 &        F9V & 30 Jan  2008 & KPNO \\
  12709 & 02 43 20.65 & +19 25 45.4 & $ 53.89\pm1.27$ & $  434.67\pm5.69$ & $  -15.50\pm5.07$ &  1214.000 &  8.21 &        dK4 &  &  \\
  12828 & 02 44 56.37 & +10 06 51.2 & $ 38.71\pm1.31$ & $  285.17\pm0.93$ & $  -30.40\pm0.98$ &  1202.200 &  4.26 &       F0IV &  &  \\
  13081 & 02 48 09.10 & +27 04 07.0 & $ 44.71\pm1.15$ & $  264.17\pm1.24$ & $ -127.75\pm0.81$ &  6127.000 &  7.60 &        K1V & 30 Jan  2008 & KPNO \\
  16846 & 03 36 47.29 & +00 35 15.9 & $ 34.52\pm0.87$ & $  -32.98\pm0.93$ & $ -163.45\pm0.88$ &  1152.000 &  8.83 &        K1V & 22 Jan  2008 & CTIO \\
  17207 & 03 41 10.52 & +03 36 40.9 & $ 39.87\pm2.01$ & $  -41.66\pm2.52$ & $ -236.97\pm1.93$ &    31.155 &  9.96 &        M0V & 23 Jan  2008 & CTIO \\
  21433 & 04 36 06.21 & +55 24 44.1 & $ 34.23\pm1.45$ & $  547.13\pm1.36$ & $ -303.46\pm1.04$ &   330.330 &  8.35 &        K2V & 31 Jan  2008 & KPNO \\
  21482 & 04 36 48.09 & +27 07 57.2 & $ 56.02\pm1.21$ & $  232.36\pm1.30$ & $ -147.11\pm1.02$ &     1.788 &  8.25 &      dK5pe & 30 Jan  2008 & KPNO \\
  21832 & 04 41 36.32 & +42 07 06.5 & $ 35.31\pm1.07$ & $  536.05\pm1.26$ & $ -416.89\pm1.09$ &  1481.000 &  7.29 &        G2V &  &  \\
  23783 & 05 06 40.66 & +51 35 53.3 & $ 38.14\pm0.79$ & $  -29.31\pm0.73$ & $ -173.95\pm0.56$ &   391.700 &  4.99 &        F0V &  &  \\
  24608 & 05 16 41.30 & +45 59 56.5 & $ 77.29\pm0.89$ & $   75.52\pm0.77$ & $ -427.13\pm0.50$ &   104.021 &  0.06 &      G5III &  &  \\
  30630 & 06 26 10.32 & +18 45 26.3 & $ 68.20\pm1.10$ & $ -119.32\pm1.06$ & $ -164.06\pm0.76$ &     6.992 &  6.79 &        dK3 & 21 Jan  2008 & CTIO \\
  32349 & 06 45 09.25 & -16 42 47.3 & $379.21\pm1.58$ & $ -546.01\pm1.33$ & $-1223.08\pm1.24$ & 18276.699 & -1.47 &        A1V &  &  \\
  34603 & 07 10 02.16 & +38 31 54.4 & $157.24\pm3.32$ & $ -439.68\pm5.33$ & $ -948.36\pm2.78$ &    10.428 & 11.48 &       dM5e & 30 Jan  2008 & KPNO \\
  34567 & 07 09 35.47 & +25 43 44.7 & $ 40.68\pm1.02$ & $ -123.09\pm0.99$ & $ -175.05\pm0.67$ &    32.807 &  7.09 &        G8V & 24 Jan  2008 & CTIO \\
  36850 & 07 34 36.00 & +31 53 19.1 & $ 63.27\pm1.23$ & $ -206.33\pm1.60$ & $ -148.18\pm1.47$ &     9.213 &  1.58 &        A1V & 31 Jan  2008 & KPNO \\
  37279 & 07 39 18.54 & +05 13 39.0 & $285.93\pm0.88$ & $ -716.57\pm0.88$ & $-1034.58\pm0.38$ & 14847.100 &  0.35 &       F5IV &  &  \\
  38382 & 07 51 46.34 & -13 53 49.9 & $ 59.98\pm0.95$ & $  -68.46\pm1.11$ & $ -344.83\pm1.03$ &  8467.000 &  5.16 &        G1V & 21 Jan  2008 & CTIO \\
  38625 & 07 54 34.10 & -01 24 44.0 & $ 52.01\pm1.85$ & $ -251.57\pm2.07$ & $  -62.07\pm1.48$ &   450.400 &  7.43 &        G8V & 22 Jan  2008 & CTIO \\
  39064 & 07 59 33.93 & +20 50 38.0 & $ 43.21\pm0.96$ & $  180.46\pm0.91$ & $ -544.36\pm0.50$ &  3138.000 &  7.70 &        K0V & 03 May  2007 & KPNO \\
  40167 & 08 12 12.73 & +17 38 52.0 & $ 39.11\pm1.38$ & $   28.29\pm2.00$ & $ -150.94\pm1.15$ &  6302.000 &  6.20 &        G5V & 23 Jan  2008 & CTIO \\
  41211 & 08 24 35.14 & -03 45 04.2 & $ 36.75\pm0.87$ & $ -211.00\pm0.82$ & $  -25.83\pm0.74$ &     1.563 &  5.59 &         F1 & 21 Jan  2008 & CTIO \\
  42172 & 08 35 51.05 & +06 37 13.9 & $ 37.68\pm1.41$ & $ -134.77\pm2.05$ & $ -131.00\pm1.42$ &    14.296 &  5.99 &         F5 & 22 Jan  2008 & CTIO \\
  43557 & 08 52 16.30 & +08 03 48.6 & $ 41.42\pm1.19$ & $  153.13\pm1.15$ & $ -235.45\pm0.66$ &    10.250 &  6.57 &        dG1 & 22 Jan  2008 & CTIO \\
  44127 & 08 59 12.84 & +48 02 32.5 & $ 68.32\pm0.79$ & $ -441.12\pm0.84$ & $ -215.21\pm0.48$ &  4028.000 &  3.14 &        A7V & 04 May  2007 & KPNO \\
  44248 & 09 00 38.75 & +41 47 00.4 & $ 60.86\pm1.30$ & $ -487.67\pm1.42$ & $ -219.29\pm0.76$ &  7980.700 &  3.97 &        F5V & 06 May  2007 & KPNO \\
  45170 & 09 12 17.55 & +14 59 45.7 & $ 48.83\pm0.92$ & $ -524.47\pm0.99$ & $  245.13\pm0.46$ &   987.920 &  6.77 &        G9V & 23 Jan  2008 & CTIO \\
  46509 & 09 29 08.90 & -02 46 08.3 & $ 58.48\pm3.80$ & $  100.93\pm4.15$ & $   -3.15\pm2.37$ &  2807.000 &  4.60 &        F6V & 07 June 2007 & CTIO \\
  48833 & 09 57 41.14 & +41 03 20.5 & $ 34.61\pm0.71$ & $ -116.94\pm0.69$ & $  -26.26\pm0.49$ &     9.284 &  5.12 &        F5V & 06 May  2007 & KPNO \\
  49809 & 10 10 05.96 & -12 48 56.4 & $ 36.61\pm0.79$ & $ -122.30\pm0.89$ & $ -109.18\pm0.53$ &    28.098 &  5.30 &        F5V & 04 June 2007 & CTIO \\
  51157 & 10 26 59.50 & +26 38 29.2 & $ 34.76\pm1.09$ & $  168.41\pm1.79$ & $  -86.32\pm0.78$ &  1180.600 &  8.24 &        K1V & 06 May  2007 & KPNO \\
  51986 & 10 37 18.26 & -48 13 32.2 & $ 37.71\pm0.51$ & $ -131.48\pm0.51$ & $   -1.58\pm0.58$ &    10.210 &  3.83 &       F3IV & 05 June 2007 & CTIO \\
  55642 & 11 23 55.37 & +10 31 46.9 & $ 41.26\pm1.16$ & $  140.75\pm1.26$ & $  -77.80\pm0.92$ & 70126.500 &  3.93 &       F2IV & 07 June 2007 & CTIO \\
  56809 & 11 38 45.39 & +45 06 30.2 & $ 42.94\pm0.95$ & $ -593.87\pm0.68$ & $   14.80\pm0.52$ &    23.541 &  8.40 &        dK5 & 03 May  2007 & KPNO \\
  56829 & 11 38 59.82 & +42 19 39.9 & $ 50.61\pm1.15$ & $ -130.44\pm0.86$ & $  436.44\pm0.73$ &    12.917 &  8.40 &        K0V & 03 May  2007 & KPNO \\
  59750 & 12 15 10.56 & -10 18 44.6 & $ 44.34\pm1.01$ & $   31.97\pm0.74$ & $-1012.44\pm0.77$ &   853.200 &  6.11 &        F5V & 04 June 2007 & CTIO \\
  63406 & 12 59 32.78 & +41 59 12.4 & $ 41.36\pm1.48$ & $ -235.87\pm0.81$ & $  181.75\pm0.99$ &   710.600 &  8.62 &        K3V & 04 May  2007 & KPNO \\
  63613 & 13 02 15.78 & -71 32 55.7 & $ 35.91\pm0.73$ & $  263.58\pm0.61$ & $  -23.28\pm0.52$ &   847.000 &  3.61 &      K2III & 05 June 2007 & CTIO \\
  64219 & 13 09 42.54 & -22 11 33.4 & $ 36.82\pm0.85$ & $  135.68\pm0.82$ & $ -339.59\pm0.56$ &    20.493 &  7.37 &        dG7 & 04 June 2007 & CTIO \\
  65378 & 13 23 55.42 & +54 55 31.5 & $ 41.73\pm0.61$ & $  121.23\pm0.48$ & $  -22.01\pm0.48$ &   175.550 &  3.95 &         Am & 04 May  2007 & KPNO \\
  67153 & 13 45 41.57 & -33 02 36.1 & $ 51.91\pm0.75$ & $ -461.85\pm0.68$ & $ -146.17\pm0.63$ &     9.945 &  4.23 &      F2III & 06 June 2007 & CTIO \\
  67927 & 13 54 41.12 & +18 23 54.9 & $ 88.17\pm0.75$ & $  -60.95\pm0.97$ & $ -358.10\pm0.82$ &   494.173 &  2.69 &       G0IV & 07 June 2007 & CTIO \\
  68682 & 14 03 32.30 & +10 47 15.1 & $ 60.24\pm0.78$ & $   85.26\pm0.64$ & $ -304.04\pm0.47$ &  3618.450 &  6.36 &        G8V & 06 June 2007 & CTIO \\
  72603 & 14 50 41.18 & -15 59 50.1 & $ 42.26\pm1.04$ & $ -135.93\pm0.94$ & $  -59.47\pm0.62$ &  5227.000 &  5.15 &        F3V & 04 June 2007 & CTIO \\
  72848 & 14 53 24.04 & +19 09 08.2 & $ 86.69\pm0.81$ & $ -442.75\pm0.57$ & $  216.84\pm0.70$ &   125.369 &  6.00 &        K1V & 07 June 2007 & CTIO \\
  73695 & 15 03 47.68 & +47 39 14.5 & $ 78.39\pm1.03$ & $ -436.24\pm1.20$ & $   18.94\pm1.17$ &     0.268 &  6.10 &        G2V & 03 May  2007 & KPNO \\
  75312 & 15 23 12.23 & +30 17 17.7 & $ 53.70\pm1.24$ & $  125.77\pm0.62$ & $ -176.48\pm0.81$ & 15179.500 &  4.98 &        G0V & 04 May  2007 & KPNO \\
  75718 & 15 28 09.61 & -09 20 53.0 & $ 50.34\pm1.11$ & $   72.69\pm1.09$ & $ -363.37\pm0.79$ &   889.620 &  6.83 &        K1V & 05 June 2007 & CTIO \\
  76267 & 15 34 41.19 & +26 42 53.7 & $ 43.65\pm0.79$ & $  120.38\pm0.55$ & $  -89.44\pm0.52$ &    17.360 &  2.23 &     B9.5IV & 06 May  2007 & KPNO \\
  77409 & 15 48 09.96 & +74 24 50.8 & $ 37.95\pm0.83$ & $  111.68\pm0.71$ & $ -304.10\pm0.90$ &   233.112 &  9.30 &         K5 & 03 May  2007 & KPNO \\
  77725 & 15 52 08.24 & +10 52 28.1 & $ 44.27\pm1.56$ & $ -267.30\pm2.17$ & $ -238.36\pm1.76$ &  1014.500 &  9.38 &        M0V & 06 June 2007 & CTIO \\
  77801 & 15 53 12.19 & +13 11 52.8 & $ 57.27\pm0.88$ & $ -150.56\pm1.32$ & $ -562.69\pm0.98$ &   138.603 &  6.10 &       G0IV & 07 June 2007 & CTIO \\
  78527 & 16 01 53.70 & +58 33 52.0 & $ 47.79\pm0.54$ & $ -320.07\pm0.50$ & $  334.96\pm0.53$ &     3.071 &  4.01 &     F8IV-V & 06 May  2007 & KPNO \\
  78709 & 16 04 03.71 & +25 15 17.4 & $ 46.56\pm0.89$ & $ -488.79\pm0.58$ & $  696.64\pm0.73$ &  4450.800 &  7.06 &        G8V & 04 May  2007 & KPNO \\
  79607 & 16 14 41.04 & +33 51 31.8 & $ 46.11\pm0.98$ & $ -266.47\pm0.86$ & $  -86.88\pm1.12$ &     1.140 &  5.64 &        G0V & 06 May  2007 & KPNO \\
  80686 & 16 28 27.80 & -70 05 04.8 & $ 82.61\pm0.57$ & $  199.89\pm0.31$ & $  110.77\pm0.51$ &    12.976 &  4.90 &        G0V & 05 June 2007 & CTIO \\
  80925 & 16 31 30.35 & -39 00 41.3 & $ 40.60\pm1.75$ & $ -428.05\pm1.47$ & $ -333.41\pm1.43$ &    31.846 &  7.25 &        dK1 & 04 June 2007 & CTIO \\
  81693 & 16 41 17.48 & +31 36 06.8 & $ 92.63\pm0.60$ & $ -462.58\pm0.59$ & $  345.05\pm0.66$ & 12596.100 &  2.80 &       G0IV & 04 May  2007 & KPNO \\
  82860 & 16 56 01.36 & +65 08 04.8 & $ 66.28\pm0.48$ & $  238.05\pm0.38$ & $   50.84\pm0.57$ &    52.109 &  4.88 &        F6V & 03 May  2007 & KPNO \\
  83608 & 17 05 20.12 & +54 28 12.2 & $ 37.08\pm0.89$ & $  -66.00\pm0.98$ & $   73.86\pm1.03$ & 42309.000 &  5.83 &        F7V & 04 May  2007 & KPNO \\
  85667 & 17 30 23.87 & -01 03 45.0 & $ 60.80\pm1.42$ & $ -126.64\pm1.72$ & $ -172.00\pm0.91$ & 16830.400 &  5.30 &     G8IV-V & 04 June 2007 & CTIO \\
  86036 & 17 34 59.59 & +61 52 28.4 & $ 70.98\pm0.55$ & $  277.38\pm0.54$ & $ -525.62\pm0.60$ & 27087.000 &  5.23 &        G1V & 06 May  2007 & KPNO \\
  86201 & 17 36 57.09 & +68 45 25.9 & $ 42.62\pm0.53$ & $    1.34\pm0.59$ & $  321.05\pm0.62$ &     5.280 &  4.80 &        F5V &  &  \\
  86400 & 17 39 16.91 & +03 33 18.9 & $ 93.36\pm1.25$ & $ -179.67\pm0.75$ & $  -98.24\pm0.52$ &    83.728 &  6.52 &        K3V & 06 June 2007 & CTIO \\
  86722 & 17 43 15.64 & +21 36 33.2 & $ 42.45\pm0.98$ & $ -123.15\pm1.00$ & $ -619.84\pm0.88$ &  2558.400 &  7.50 &        K0V & 07 June 2007 & CTIO \\
  88601 & 18 05 27.21 & +02 30 08.8 & $196.62\pm1.38$ & $  124.56\pm1.15$ & $ -962.66\pm0.91$ & 32188.801 &  4.02 &        K0V & 06 June 2007 & CTIO \\
  89937 & 18 21 02.34 & +72 44 01.3 & $124.11\pm0.48$ & $  531.08\pm0.49$ & $ -351.59\pm0.46$ &   280.550 &  3.57 &        F7V &  &  \\
  90355 & 18 26 10.00 & +08 46 39.0 & $ 37.04\pm1.70$ & $ -194.69\pm1.45$ & $ -458.10\pm1.27$ &   293.500 &  7.78 &        G7V & 05 June 2007 & CTIO \\
  91009 & 18 33 55.60 & +51 43 11.7 & $ 60.90\pm0.73$ & $  186.62\pm0.71$ & $ -324.89\pm0.77$ &     5.975 &  8.04 &       dM0e &  &  \\
  92919 & 18 55 53.14 & +23 33 26.4 & $ 46.64\pm1.03$ & $  130.79\pm0.59$ & $ -283.07\pm0.74$ &     2.879 &  8.30 &        K0V & 07 June 2007 & CTIO \\
  93017 & 18 57 01.61 & +32 54 04.6 & $ 66.76\pm0.54$ & $  202.85\pm0.46$ & $ -143.97\pm0.49$ & 22423.000 &  5.22 &        G0V &  &  \\
  93174 & 18 58 43.47 & -37 06 25.5 & $ 33.43\pm0.92$ & $ -132.25\pm1.36$ & $ -110.45\pm0.77$ &     0.591 &  4.74 &        F0V & 04 June 2007 & CTIO \\
  93926 & 19 07 32.39 & +30 15 16.2 & $ 35.70\pm0.78$ & $  111.96\pm0.63$ & $  103.03\pm0.68$ &     2.131 &  7.63 &        G8V &  &  \\
  93966 & 19 07 57.28 & +16 51 14.9 & $ 47.72\pm0.77$ & $   65.61\pm0.67$ & $ -304.46\pm0.60$ &    21.947 &  6.08 &        G5V & 06 June 2007 & CTIO \\
  95575 & 19 26 25.98 & +49 27 55.1 & $ 39.73\pm1.03$ & $  457.19\pm1.09$ & $  713.97\pm1.14$ &   166.360 &  8.01 &        K3V &  &  \\
  95995 & 19 31 08.54 & +58 35 13.1 & $ 59.84\pm0.64$ & $ -510.04\pm0.68$ & $ -397.54\pm0.68$ &   491.900 &  6.59 &        K1V &  &  \\
  97944 & 19 54 17.82 & -23 56 24.3 & $ 70.34\pm0.81$ & $ -122.67\pm0.78$ & $ -409.86\pm0.53$ &    46.817 &  6.16 &        K3V & 04 June 2007 & CTIO \\
  98416 & 19 59 47.49 & -09 57 26.2 & $ 40.75\pm1.35$ & $ -246.73\pm2.31$ & $ -392.36\pm1.63$ &  1786.270 &  6.22 &        F8V & 06 June 2007 & CTIO \\
 101382 & 20 32 51.76 & +41 53 50.6 & $ 44.99\pm0.64$ & $ -156.89\pm0.53$ & $  452.80\pm0.47$ &    57.321 &  7.09 &        G9V &  &  \\
 101750 & 20 37 20.82 & +75 35 56.7 & $ 36.16\pm0.97$ & $  309.20\pm1.12$ & $  539.49\pm0.97$ &     0.278 &  7.23 &        K0V &  &  \\
 101769 & 20 37 32.87 & +14 35 42.7 & $ 33.49\pm0.88$ & $  118.28\pm1.33$ & $  -47.65\pm1.09$ &  9733.700 &  3.62 &       F5IV & 07 June 2007 & CTIO \\
 101955 & 20 39 37.20 & +04 58 18.7 & $ 53.82\pm2.21$ & $  862.35\pm2.15$ & $   67.57\pm1.78$ &   920.200 &  7.89 &        K5V & 05 June 2007 & CTIO \\
 102431 & 20 45 21.12 & +57 34 47.0 & $ 36.87\pm0.46$ & $  -62.95\pm0.41$ & $ -235.56\pm0.45$ &   523.360 &  4.51 &       F8IV &  &  \\
 103655 & 21 00 05.35 & +40 04 13.0 & $ 66.21\pm2.54$ & $  614.37\pm2.07$ & $ -247.16\pm2.45$ & 10777.100 & 10.10 &       dM3e &  &  \\
 104858 & 21 14 28.79 & +10 00 27.8 & $ 54.11\pm0.85$ & $   42.32\pm0.89$ & $ -303.43\pm0.57$ &  2082.100 &  4.49 &        F5V & 05 June 2007 & CTIO \\
 105312 & 21 19 46.00 & -26 21 07.2 & $ 53.40\pm1.09$ & $ -582.35\pm1.11$ & $ -357.67\pm0.62$ &    21.346 &  6.55 &        G5V & 04 June 2007 & CTIO \\
 105406 & 21 21 01.44 & +40 20 44.1 & $ 37.64\pm0.59$ & $  -18.89\pm0.43$ & $ -208.92\pm0.47$ &     3.243 &  6.43 &        F8V &  &  \\
 107089 & 21 41 28.47 & -77 23 22.1 & $ 47.22\pm2.90$ & $   64.83\pm2.63$ & $ -240.38\pm2.12$ &  1020.000 &  3.75 &      K0III & 06 June 2007 & CTIO \\
 107556 & 21 47 02.29 & -16 07 35.6 & $ 84.58\pm0.88$ & $  263.26\pm1.23$ & $ -296.23\pm0.67$ &     1.023 &  2.83 &         Am & 04 June 2007 & CTIO \\
 109176 & 22 07 00.47 & +25 20 42.2 & $ 85.06\pm0.71$ & $  296.73\pm0.56$ & $   26.93\pm0.69$ &    10.213 &  3.76 &        F5V &  &  \\
 111170 & 22 31 18.22 & -06 33 17.6 & $ 39.18\pm1.83$ & $  160.65\pm1.15$ & $ -108.63\pm0.80$ &   630.140 &  6.27 &        F7V &  &  \\
 111802 & 22 38 45.29 & -20 37 15.4 & $115.71\pm1.50$ & $  450.58\pm1.61$ & $  -79.86\pm1.15$ &     4.083 &  9.09 &       dM1e & 07 June 2007 & CTIO \\
 113718 & 23 01 51.54 & -03 50 55.4 & $ 59.04\pm3.40$ & $  395.50\pm1.21$ & $ -207.11\pm1.17$ &   454.660 &  7.46 &        K4V &  &  \\
 113860 & 23 03 29.76 & -34 44 58.6 & $ 34.98\pm0.80$ & $   74.80\pm0.81$ & $   84.45\pm0.70$ &   178.318 &  5.10 &        F0V & 07 June 2007 & CTIO \\
 114379 & 23 09 57.23 & +47 57 30.0 & $ 39.56\pm7.67$ & $  147.06\pm5.75$ & $   12.42\pm6.72$ &     3.033 &  7.91 &        K2V &  &  \\
 115126 & 23 19 06.51 & -13 27 30.4 & $ 48.22\pm5.25$ & $  265.64\pm6.17$ & $  -44.87\pm5.22$ &  2323.600 &  5.21 &       G5IV &  &  \\
 116584 & 23 37 33.71 & +46 27 33.0 & $ 38.74\pm0.68$ & $  159.22\pm0.33$ & $ -421.46\pm0.51$ &    20.521 &  3.88 &   G8III-IV &  &  \\
 116727 & 23 39 20.85 & +77 37 56.2 & $ 72.50\pm0.52$ & $  -48.85\pm0.48$ & $  127.18\pm0.44$ & 24135.000 &  3.23 &       K1IV &  &  \\
 117712 & 23 52 24.52 & +75 32 40.2 & $ 92.68\pm0.55$ & $  341.82\pm0.53$ & $   41.88\pm0.47$ &     7.753 &  6.40 &        K3V &  &  \\

\enddata
\tablenotetext{a}{Data in these columns derived from the Hipparcos database \citep{hip}.}
\tablenotetext{b}{Data in these columns derived from the Ninth Catalog of Spectroscopic Binaries \citep{sb9}.}
\tablenotetext{c}{Spectral Types taken from those given in the Ninth Catalog of Spectroscopic Binaries \citep{sb9}.}
\label{tab:targs}

\end{deluxetable}

\begin{deluxetable}{cccccccccccccc}
\tabletypesize{\scriptsize}
\rotate
\tablewidth{0pt}
\tablecaption{Candidate and Known CPM Companions}
\tablehead{
\colhead{Primary} & \colhead{2MASS RA} & \colhead{2MASS DEC} & \colhead{$\mu_{\alpha}$\tablenotemark{a}} & \colhead{$\mu_{\delta}$\tablenotemark{a}} & \colhead{Sep.} & \colhead{$M_J$\tablenotemark{b}} & \colhead{$J-K_s$\tablenotemark{c}} & \colhead{$J-H$\tablenotemark{c}} & \colhead{$H-K_s$\tablenotemark{c}} & \colhead{SpT\tablenotemark{d}} & \colhead{SpT} & \colhead{Notes} & \colhead{Companion}\\
\colhead{Hip.\ $\#$} & \colhead{Tertiary} & \colhead{Tertiary} & \colhead{(mas/yr)} & \colhead{(mas/yr)} & \colhead{(AU)} & \colhead{(mag)} & \colhead{(mag)} & \colhead{(mag)} & \colhead{(mag)} & \colhead{Tert.} & \colhead{Pri.} & \colhead{} & \colhead{Status} }

\startdata

 13081 & 02h48m09.7s & $+$27d04m25s &  $237.1\pm40$ &  $-122.1\pm44$ &  460    &  $8.980\pm0.022$ &  $0.862\pm0.031$ &  $0.554\pm0.040$ &  $0.308\pm0.040$ & M4.5    & K1V   & 1\tablenotemark{e} & Yes \\
 16846 & 03h36m46.8s & $+$00d35m15s &  $-28.2\pm32$ &  $-131.6\pm32$ &  220    &          \nodata &          \nodata &\nodata           &\nodata           & K6      & K1V   & 5\tablenotemark{f} & Yes \\
 21482 & 04h36m44.9s & $+$27d09m51s &  $254.1\pm44$ &  $-201.7\pm40$ & 2180    & $13.340\pm0.038$ &  $0.462\pm0.079$ &  $0.366\pm0.069$ &  $0.096\pm0.090$ & DC      & dK5pe & 1\tablenotemark{g} & Yes \\
 36850 & 07h34m37.4s & $+$31d52m10s & $-207.6\pm28$ &   $-96.0\pm28$ & 1130    &  $5.079\pm0.018$ &  $0.837\pm0.027$ &  $0.653\pm0.028$ &  $0.184\pm0.029$ & M0.5    & A1V   & 5\tablenotemark{h} & Yes \\
 41211 & 08h24m33.8s & $-$03d44m34s & $-211.5\pm36$ &   $-14.2\pm36$ &  980    &  $9.267\pm0.027$ &  $0.841\pm0.037$ &  $0.547\pm0.035$ &  $0.294\pm0.034$ & M4.5    & F1V   & 2\tablenotemark{f} & Yes \\
 41211 & 08h24m52.3s & $-$03d41m02s & $-217.0\pm36$ &   $-17.0\pm36$ & 9640    &  $9.374\pm0.026$ &  $0.918\pm0.032$ &  $0.566\pm0.033$ &  $0.352\pm0.028$ & M5.5    & F1V   & 1\tablenotemark{i} & Yes \\
 42172 & 08h35m51.3s & $+$06d37m22s & $-120.7\pm48$ &  $-133.3\pm44$ &  230    &  $4.033\pm0.024$ &  $0.380\pm0.032$ &  $0.170\pm0.036$ &  $0.110\pm0.034$ & G5      & F5V   & 5\tablenotemark{j} & Yes \\
 45170 & 09h12m14.7s & $+$14d59m40s & $-565.5\pm32$ &   $279.3\pm40$ &  860    & $13.955\pm0.078$ &  $1.469\pm0.100$ &  $0.891\pm0.122$ &  $0.578\pm0.103$ & L8      & G9V   & 1\tablenotemark{k} & Yes \\
 46509 & 09h29m09.2s & $-$02d45m03s &  $138.7\pm36$ &   $-20.6\pm40$ & 1120    &  $4.529\pm0.029$ &  $0.557\pm0.036$ &  $0.391\pm0.057$ &  $0.166\pm0.054$ & K0      & F6V   & 5\tablenotemark{j} & Yes \\
 72603 & 14h50m27.4s & $-$16d03m23s & $-112.5\pm28$ &  $-105.6\pm32$ & \nodata & $14.603\pm0.120$ &  $1.065\pm0.257$ &  $0.535\pm0.206$ &  $0.530\pm0.281$ & \nodata & F3V   & 3 & No  \\
 75312 & 15h23m22.6s & $+$30d14m56s &  $133.1\pm32$ &  $-183.2\pm28$ & 3640    & $14.706\pm0.099$ &  $1.708\pm0.120$ &  $1.128\pm0.128$ &  $0.580\pm0.105$ & L8      & G0V   & 1\tablenotemark{l} & Yes \\
 75718 & 15h28m12.2s & $-$09d21m28s &   $83.5\pm28$ &  $-355.9\pm28$ &  980    &  $4.504\pm0.027$ &  $0.533\pm0.034$ &  $0.443\pm0.048$ &  $0.090\pm0.045$ & K2      & K1V   & 5\tablenotemark{g} & Yes \\
 83608 & 17h05m20.3s & $+$54d28m00s &  $-74.5\pm24$ &    $89.5\pm36$ &  340    &  $6.866\pm0.043$ &  $0.591\pm0.046$ &  $0.356\pm0.051$ &  $0.235\pm0.032$ & ??      & F7V   & 2\tablenotemark{m} & Yes \\
 86722 & 17h43m15.4s & $+$21d36m11s &  $-98.7\pm40$ &  $-631.5\pm40$ &  530    &  $9.650\pm0.025$ &  $0.811\pm0.031$ &  $0.495\pm0.033$ &  $0.316\pm0.028$ & M4.5    & K0V   & 3 & Yes \\
 92919 & 18h55m50.4s & $+$23d36m51s &  $-29.0\pm28$ &  $-154.9\pm28$ & \nodata & $10.797\pm0.021$ &  $0.816\pm0.029$ &  $0.579\pm0.030$ &  $0.237\pm0.029$ & \nodata & K0V   & 4 & No  \\
 97944 & 19h54m20.6s & $-$23d56m40s & $-151.7\pm32$ &  $-400.8\pm32$ &  590    &  $9.027\pm0.021$ &  $0.952\pm0.031$ &  $0.598\pm0.032$ &  $0.354\pm0.033$ & M5      & K3V   & 3 & Yes \\
101769 & 20h37m33.7s & $+$14d35m32s &   $50.9\pm36$ &   $-55.6\pm36$ & \nodata &  $8.086\pm0.046$ & $-0.008\pm0.096$ &  $0.358\pm0.070$ & $-0.366\pm0.099$ & G0      & F5IV  & 3 & No  \\
111802 & 22h38m45.3s & $-$20d36m52s &  $443.1\pm40$ &   $-25.2\pm36$ &  200    &  $7.661\pm0.024$ &  $0.853\pm0.029$ &  $0.527\pm0.062$ &  $0.326\pm0.059$ & M3.5    & M1.5V & 1\tablenotemark{e} & Yes \\

\enddata

\tablenotetext{a}{Listed proper motion measurements are derived from the data presented in this paper save for those objects with a [5] in the follow-up column.  Those values were taken from Hipparcos, as those objects were saturated in our data.}
\tablenotetext{b}{Absolute J magnitudes derived from 2MASS photometry and Hipparcos parallaxes.}
\tablenotetext{c}{All photometry listed is from the 2MASS PSC.}
\tablenotetext{d}{Companion spectral types were based either on SpeX prism spectroscopy obtained for this work (objects with [2] or [3] in the Notes column), or were already known in which case the spectral type listed is from the paper referenced by the lettered superscripts in the Notes column (objects with [1] or [5] in the Notes column).}
\tablenotetext{e}{\citet{inr95}}
\tablenotetext{f}{\citet{tok06}}
\tablenotetext{g}{\citet{mzh08}}
\tablenotetext{h}{\citet{js1926}}
\tablenotetext{i}{\citet{inr03}}
\tablenotetext{j}{\citet{close1990}}
\tablenotetext{k}{\citet{wil00}}
\tablenotetext{l}{\citet{gizis01}}
\tablenotetext{m}{\citet{ccdm2}}
\tablecomments{(1) Previously known companion, no additional follow-up data required. (2) Previously known candidate companion, no Spectral Type, SpeX data acquired. (3) No previous data, SpeX data acquired. (4) Additional photometry acquired from Vizier database, not a companion.  (5) Previously known bright companion.  Saturated in these data, found via literature search using the SIMBAD database.}
\label{tab:cand}

\end{deluxetable}

\begin{deluxetable}{cccc}
\tabletypesize{\footnotesize}
\tablewidth{0pt}
\tablecaption{Companion Fraction by Spectral Type}
\tablehead{
\colhead{Spectral} & \colhead{Number of} & \colhead{Number with} & \colhead{Companion}\\
\colhead{Type} & \colhead{Primaries} & \colhead{Tertiaries} & \colhead{Fraction}}

\startdata

B &  1 & 0 & $ 0.0^{+60.1}_{ -0.0}\%$ \\
A &  4 & 1 & $25.0^{+27.4}_{-10.3}\%$ \\
F & 23 & 5 & $21.7^{+10.7}_{ -6.1}\%$ \\
G & 21 & 2 & $ 9.5^{+10.3}_{ -3.1}\%$ \\
K & 24 & 6 & $25.0^{+10.5}_{ -6.6}\%$ \\
M &  4 & 1 & $25.0^{+27.4}_{-10.3}\%$ \\

\enddata
\label{tab:sptcf}
\end{deluxetable}

\begin{deluxetable}{cccc}
\tabletypesize{\footnotesize}
\tablewidth{0pt}
\tablecaption{Companion Fraction by Spectroscopic Binary Period}
\tablehead{
\colhead{Period} & \colhead{Number of} & \colhead{Number with} & \colhead{Companion}\\
\colhead{{Log Days}} & \colhead{Primaries} & \colhead{Tertiaries} & \colhead{Fraction}}

\startdata

-0.5 &  2 & 0 & $ 0.0^{+45.8}_{ -0.0}\%$ \\
 0.5 & 12 & 6 & $50.0^{+13.3}_{-13.3}\%$ \\
 1.5 & 19 & 1 & $ 5.3^{+10.2}_{ -1.7}\%$ \\
 2.5 & 18 & 2 & $11.1^{+11.6}_{ -3.7}\%$ \\
 3.5 & 19 & 6 & $31.6^{+11.9}_{ -8.4}\%$ \\
 4.5 &  7 & 2 & $28.6^{+20.3}_{-10.7}\%$ \\

\enddata
\label{tab:percf}
\end{deluxetable}

\end{document}